\documentclass[twocolumn]{aastex61}
\pdfoutput=1
\usepackage{amsmath}
\usepackage{url}

\newcommand{\msun} {$M_\odot$}
\newcommand{\logg} {\log \textsl{\textrm{g}}}
\newcommand{\Teff} {T_{\rm eff}}
\newcommand\gta{\lower 0.5ex\hbox{$\buildrel > \over \sim\ $}}
\newcommand\lta{\lower 0.5ex\hbox{$\buildrel < \over \sim\ $}}

\begin{document}

\title{On the Spectral Evolution of Hot White Dwarf Stars. \\ I. A Detailed Model-Atmosphere Analysis of Hot White Dwarfs from SDSS DR12}

\author{A. B\'edard, P. Bergeron, P. Brassard, and G. Fontaine}
\affiliation{\centering D\'epartement de Physique, Universit\'e de Montr\'eal, Montr\'eal, QC H3C 3J7, Canada \\ bedard@astro.umontreal.ca, bergeron@astro.umontreal.ca, brassard@astro.umontreal.ca \vspace*{5mm}}

\shorttitle{Analysis of Hot White Dwarfs from SDSS}
\shortauthors{B\'edard, Bergeron, Brassard \& Fontaine}

\begin{abstract}
As they evolve, white dwarfs undergo major changes in surface composition, a phenomenon known as spectral evolution. In particular, some stars enter the cooling sequence with helium atmospheres (type DO) but eventually develop hydrogen atmospheres (type DA), most likely through the upward diffusion of residual hydrogen. Our empirical knowledge of this process remains scarce: the fractions of white dwarfs that are born helium-rich and that experience the DO-to-DA transformation are poorly constrained. We tackle this issue by performing a detailed model-atmosphere investigation of 1806 hot ($\Teff \ge 30,000$ K) white dwarfs observed spectroscopically by the Sloan Digital Sky Survey. We first introduce our new generations of model atmospheres and theoretical cooling tracks, both appropriate for hot white dwarfs. We then present our spectroscopic analysis, from which we determine the atmospheric and stellar parameters of our sample objects. We find that $\sim$24\% of white dwarfs begin their degenerate life as DO stars, among which $\sim$2/3 later become DA stars. We also infer that the DO-to-DA transition occurs at substantially different temperatures ($75,000 \ {\rm K} > \Teff > 30,000$ K) for different objects, implying a broad range of hydrogen content within the DO population. Furthermore, we identify 127 hybrid white dwarfs, including 31 showing evidence of chemical stratification, and we discuss how these stars fit in our understanding of the spectral evolution. Finally, we uncover significant problems in the spectroscopic mass scale of very hot ($\Teff > 60,000$ K) white dwarfs.
\end{abstract}

\section{Introduction} \label{sec:intro}

White dwarf stars form a remarkably homogeneous class of objects, and yet they exhibit a surprisingly diverse variety of faces. On one hand, all stars with initial masses lower than $\sim$8 \msun\ (that is, $\sim$97\% of the stars in the Galaxy) are destined to evolve into white dwarfs, all sharing the same fundamental, defining property: an exceptionally high density, reaching up to $\sim$10$^6-10^8$ g cm$^{-3}$ in the stellar core \citep{fontaine2001}. Indeed, following the exhaustion of nuclear energy sources, gravitational contraction proceeds unimpeded until it is eventually halted by electron degeneracy pressure, thus producing a highly compact stellar remnant, condemned to simply cool off with time. Such an object is characterized by an extreme surface gravity field of typically $\sim$10$^8$ cm s$^{-2}$, in which the process of gravitational settling is predicted to be so efficient that all heavy elements should sink into the star on very short timescales, leaving only the lightest element at the surface \citep{paquette1986,dupuis1992}. Consequently, all white dwarfs are expected to have a stratified chemical structure: a core made of carbon and oxygen, surrounded by a thin pure-helium envelope, itself surrounded by an even thinner pure-hydrogen layer encompassing the observable atmosphere \citep{fontaine2001}.

On the other hand, while this simple picture provides a fairly accurate description of reality, it does not reflect the numerous details uncovered by decades of observational and theoretical investigations. It is well-known that although most white dwarfs have canonical hydrogen-rich atmospheres, a significant fraction of them instead have helium-rich atmospheres. Furthermore, in several instances, small traces of elements other than the main constituent are detected at the surface. Even more surprising is the fact that the atmospheric composition of a star can change radically as it evolves on the cooling sequence, a phenomenon referred to as the spectral evolution of white dwarfs. This empirical evidence indicates that various transport mechanisms, such as ordinary diffusion, thermal diffusion, radiative levitation, convective mixing, convective overshooting, stellar winds, and accretion, compete with gravitational settling in determining the chemical appearance of degenerate stars. 

More than 30 years ago, \citet{fontaine1987} laid the foundations of this most interesting field of research aiming at providing a complete, consistent theory of the spectral evolution, which would identify the predominant physical processes and explain the observed chemical peculiarities. Such an endeavor is of course crucial to our understanding of white dwarf evolution, but also of stellar evolution in general, to which white dwarfs, as the ultimate fate of the vast majority of stars, hold important clues. We now review the progress that has been made over the years toward the achievement of this goal. Since it would be impossible to cover the whole subject in detail, in the present paper we restrict our attention to the early phases of degenerate evolution, more specifically to hot white dwarfs with effective temperatures $\Teff \ge 30,000$ K. The reader is referred to \citet{rolland2018,rolland2020}, \citet{ourique2019,ourique2020}, \citet{blouin2019}, \citet{genest-beaulieu2019b}, \citet{coutu2019}, and \citet{cunningham2020} for recent studies of the spectral evolution of cooler remnants.

In the 1980s, white dwarf research was stimulated by the advent of the Palomar-Green (PG) survey \citep{green1986}, which revealed a few intriguing features. First, at the very hot end of the cooling sequence ($\Teff > 80,000$ K), a lack of hydrogen-atmosphere white dwarfs of spectral type DA, with respect to their helium-atmosphere counterparts of spectral type DO, was discovered \citep{fleming1986}. Second, it was found that the helium-dominated white dwarf sequence was interrupted by the so-called DB gap in the range $45,000 \ {\rm K} > \Teff > 30,000$ K, between the hot DO stars and cooler DB stars, where all objects had hydrogen-dominated atmospheres \citep{wesemael1985,liebert1986}. Between the former and latter regimes, the ratio of hydrogen-rich to helium-rich white dwarfs showed a monotonic rise with decreasing effective temperature.

In light of these results, \citet{fontaine1987} devised a model of the spectral evolution based on the unifying paradigm that all white dwarfs shared a common origin. In their scenario, the extremely hot hydrogen-deficient PG 1159 stars were the progenitors of the entire white dwarf population, which naturally explained the paucity of DA stars with $\Teff > 80,000$ K. It was already believed at that time that PG 1159 stars were the product of a late helium-shell flash, during which the superficial hydrogen inherited from previous evolutionary phases had been ingested in the stellar envelope and burned \citep{iben1983}. However, \citet{fontaine1987} postulated that a small amount of hydrogen had survived this violent episode, while remaining thoroughly mixed in the envelope and thus invisible. They suggested that this residual hydrogen then gradually diffused upward under the influence of gravitational settling, building up a thicker and thicker hydrogen layer at the surface, and ultimately transforming all helium-atmosphere white dwarfs into hydrogen-atmosphere white dwarfs. This idea accounted for both the increase of the DA-to-DO ratio along the cooling sequence and the existence of the blue edge of the DB gap at $\Teff \sim 45,000$ K. Finally, the reappearance of DB stars beyond the red edge of the DB gap at $\Teff \sim 30,000$ K was interpreted as the consequence of the dilution of the hydrogen layer by the growing convection zone in the underlying helium envelope.

One striking implication of this so-called float-up model was that the mass of the hydrogen layer in DA white dwarfs had to be extremely small, of the order of $q_{\rm H} \equiv M_{\rm H}/M_{\star} \sim 10^{-16}-10^{-10}$ \citep{fontaine1987}, many orders of magnitude smaller than the standard value of $10^{-4}$ predicted by evolutionary calculations \citep{iben1984}. Nevertheless, further evidence in favor of such thin hydrogen layers was obtained independently from spectral studies of hot DA stars. Indeed, a number of these objects were found to exhibit X-ray and extreme-ultraviolet (EUV) flux deficiencies with respect to what was expected from pure-hydrogen atmospheres (\citealt{vennes1988}, and references therein). The additional short-wavelength opacity was first assumed to be provided by minute traces of helium, supported in the outer layers by radiative levitation. This hypothesis was motivated by the detection of helium in the optical spectra of some hot hydrogen-rich white dwarfs, mostly through a weak He {\sc ii} $\lambda$4686 feature, defining the spectral class DAO. However, \citet{vennes1988} demonstrated that the amount of helium supported by radiation pressure was insufficient to explain the measured flux deficiencies. Instead, they showed that the short-wavelength observations could be successfully reproduced by a model in which a very thin hydrogen layer ($q_{\rm H} \sim 10^{-15}-10^{-13}$) floated on top of the helium envelope in diffusive equilibrium. In this context, DAO stars could naturally be interpreted as transitional objects currently undergoing the DO-to-DA transformation through the upward diffusion of hydrogen envisioned by \citet{fontaine1987}.

In the following years, several challenges to the float-up model emerged. Analyses of far-ultraviolet (FUV) spectra of hot white dwarfs conclusively revealed the presence of traces of numerous heavy elements (mainly carbon, nitrogen, oxygen, silicon, iron, and nickel) in the atmospheres of these objects \citep{vennes1992,holberg1993,werner1994,werner1995b}. It then became quite clear that metals, rather than helium, were responsible for the X-ray and EUV opacity in hot DA stars, and thin hydrogen layers were therefore no longer required. Another major discovery was made by \citet{napiwotzki1995} through their survey of old central stars of planetary nebulae. They identified a large number of hydrogen-atmosphere objects with $\Teff > 80,000$ K, implying that a significant proportion of the white dwarf population was born hydrogen-rich, and thus dismissing the assertion of \citet{fontaine1987} that all white dwarfs descended from PG 1159 stars. The two latter findings pointed to the existence of an evolutionary channel involving white dwarfs characterized by canonical thick hydrogen layers ($q_{\rm H} \sim 10^{-4}$) and hence retaining their DA spectral type throughout their entire evolution. Nonetheless, the idea that DO stars had to turn into DA stars with thin hydrogen layers in the DB gap, perhaps via the DAO stage, remained relevant.

In that respect, of particular importance was the detailed investigation of the float-up model conducted by \citet{macdonald1991}, who studied the combined effect of gravitational settling, ordinary diffusion, radiative levitation, and convective mixing in hydrogen-helium white dwarf envelopes. They concluded that the float-up hypothesis effectively accounted for the key observational features uncovered by the PG survey for hydrogen layer masses of $q_{\rm H} \sim 10^{-15}-10^{-13}$, but they also encountered a few inconsistencies. They noted that the amount of hydrogen had to be much smaller in DO white dwarfs than in DAO white dwarfs, without any known intermediate object in between, suggesting that these two spectral groups did not actually share an evolutionary link. In addition, they found that the hydrogen abundances measured in the cooler DBA stars were much higher than those predicted by the convective dilution scenario. These two results lead them to invoke accretion of hydrogen from the interstellar medium as the main driving force behind the spectral evolution of white dwarfs.

Another decisive step was taken by \citet{bergeron1994}, who carried out an in-depth model-atmosphere analysis of a sample of hot DA and DAO stars. In particular, they analyzed their DAO white dwarfs with two sets of models differing by the assumed chemical structure: a homogeneous configuration, with hydrogen and helium uniformly distributed throughout the atmosphere, and a stratified configuration, with hydrogen floating on top of helium in diffusive equilibrium. They showed that the optical spectra of all objects but one (PG 1305$-$017) were better reproduced by homogeneous models rather than stratified models. Moreover, for most DAO stars, they obtained effective temperatures considerably higher than the hot boundary of the DB gap. This outcome confirmed that the DAO phenomenon could not be viewed as a transitional stage between the DO and DA stars resulting from the upward diffusion of hydrogen. Nevertheless, one DAO white dwarf, PG 1305$-$017, remained consistent with this picture, since it lied precisely at the blue edge of the DB gap and displayed convincing signs of chemical stratification. \citet{bergeron1994} also noted that the bulk of their DAO sample was afflicted by the so-called Balmer-line problem, the lower lines of the Balmer series being predicted too shallow by their models. They argued that atmospheric pollution by heavy elements, and the neglect of the corresponding opacities in the model-atmosphere calculations, was the most probable cause of this issue, which was later corroborated by \citet{werner1996a} and \citet{gianninas2010}. Consequently, they proposed that a weak stellar wind, possibly driven by metals, was responsible for maintaining helium at the surface of DAO white dwarfs, as well as homogenizing the composition of the external layers.

In subsequent years, overwhelming evidence in favor of the occurrence of winds in some of the hottest ($\Teff \ \gta 60,000$ K) white dwarfs accumulated. \citet{napiwotzki1999} found that DAO stars followed a correlation of decreasing helium abundance with decreasing luminosity, which was indeed expected if the helium content was sustained by a steadily fading wind. Additionally, while it was initially believed that heavy elements were radiatively levitated in the atmospheres of hot DA and DO stars, the elaborate theoretical work of \citet{chayer1995} and \citet{dreizler1999} compellingly demonstrated that the observed abundance patterns were poorly reproduced under the assumption of an equilibrium between gravitational and radiative forces. As more and more metals (including elements beyond the iron group) were discovered and improved abundance determinations were made available, this disagreement became even more evident \citep{dreizler1996,barstow2003a,barstow2014,chayer2005,good2005,vennes2006,werner2007,werner2012}. In most cases, mass loss was put forward as the most plausible alternative mechanism (with the recent exception of \citealt{barstow2014}, who instead invoked an interplay between accretion of planetary debris and radiative levitation). Finally, \citet{werner1995a} and \citet{dreizler1995} exposed the puzzling phenomenon of the ultrahigh-excitation absorption lines in the optical spectra of a few hot helium-rich objects. Because the formation of these peculiar features, arising from very high ionization stages of light metals, required excessively high temperatures, a photospheric origin was definitely ruled out. Based on the asymmetric shape and the slight blueshift of some ultrahigh-excitation lines, they speculated that such features were generated in a shock-heated circumstellar wind, a conjecture validated only recently by \citet{reindl2019}.

The wind interpretation was put on more quantitative grounds by \citet{unglaub1998,unglaub2000}, who carried out time-dependent simulations of diffusion and mass loss in white dwarf envelopes. Adopting a metal-driven wind model, they calculated that mass-loss rates of the order of $\dot{M} \sim 10^{-12}-10^{-11}$ \msun\ yr$^{-1}$ effectively gave rise to helium pollution of hydrogen-rich envelopes with surface abundances similar to those measured in DAO stars. This support mechanism was shown to be efficient only above a certain line in the $\logg - \Teff$ diagram called the wind limit, beyond which mass loss vanished and gravitational settling rapidly transformed DAO stars into DA stars. These results underlined the essential role played by stellar winds in the DAO phenomenon. However, a significant discrepancy remained between the theoretical and empirical wind limits, found at $\Teff \sim 85,000$ and 60,000 K, respectively, for normal-mass objects \citep{gianninas2010}. In addition, \citet{unglaub2000} demonstrated that mass loss was mandatory to explain the presence of large amounts of carbon and oxygen in the atmospheres of PG 1159 stars, a conclusion also reached by \citet{quirion2012} from their investigation of the GW Vir instability strip.

The next major developments in our understanding of the spectral evolution were made possible by the Sloan Digital Sky Survey (SDSS; \citealt{york2000}), which revolutionized the white dwarf field by increasing the number of spectroscopically identified white dwarfs more than tenfold \citep{eisenstein2006b,kleinman2013,kepler2015,kepler2016,kepler2019}. On the basis of a model-atmosphere analysis of the Data Release (DR) 4 sample, \citet{eisenstein2006a} reported the striking discovery of a small number of helium-rich objects in the DB gap, indicating that not all DO stars inevitably evolved into DA stars. In other words, these hot DB stars had to be the outcome of a different evolutionary channel, in which white dwarfs were born almost completely devoid of hydrogen and thus permanently preserved helium-dominated atmospheres. Nevertheless, \citet{eisenstein2006a} estimated that a substantial deficit of helium-rich stars persisted in the range $45,000 \ {\rm K} > \Teff > 20,000$ K with respect to higher and lower effective temperatures, suggesting that such hydrogen-free evolution occurred quite rarely. Therefore, the notion of a DB gap was replaced by that of a DB deficiency, whose cool boundary was $\sim$10,000 K cooler than previously believed, a result also obtained by \citet{bergeron2011} from their assessment of the PG luminosity function.

Finally, \citet{manseau2016} recently took advantage of the large SDSS DR12 sample to search for new hybrid white dwarfs (exhibiting traces of both hydrogen and helium) with chemically layered atmospheres, similar to the unique such star PG 1305$-$017 known at the time. They successfully identified about a dozen objects that were better modeled assuming stratified atmospheres, with typical hydrogen masses of $q_{\rm H} \sim 10^{-17}-10^{-16}$. Considering that the derived effective temperatures fell in the range $55,000 \ {\rm K} > \Teff > 40,000$ K, coinciding with the onset of the DB deficiency, this finding proved that the float-up of hydrogen and the corresponding DO-to-DA transition indeed occurred for an important fraction of all hot helium-rich white dwarfs. Furthermore, \citet{manseau2016} stressed the fact that their new stratified candidates, of rather diverse spectral types (DAB, DAO, DBA, or DOA), were well separated from the classical, hotter, chemically homogeneous DAO stars in the $\logg - \Teff$ diagram, reflecting the fundamentally different physical processes at work in the two groups of objects.

This completes our historical overview of the state of knowledge regarding the spectral evolution of hot white dwarfs. Despite the latest advances, we wish to point out that the extensive spectroscopic dataset provided by the SDSS has never been fully exploited in terms of the subject at hand. In fact, several model-atmosphere studies of the hot white dwarfs in the SDSS have been published \citep{hugelmeyer2005,hugelmeyer2006,eisenstein2006b,eisenstein2006a,tremblay2011,kleinman2013,werner2014,reindl2014b,kepler2015,kepler2016,kepler2019,manseau2016,genest-beaulieu2019a}, but they all suffer from either one of the two following limitations. First, most of them do not include all relevant spectral classes (DA, DB, DO, plus the various subtypes) simultaneously, rendering it difficult to establish a global picture of the spectral evolution. Second, the few of them that do consider all spectral types are geared toward classification purposes and hence provide only coarse atmospheric parameter estimates. In particular, they make use of model atmospheres assuming local thermodynamic equilibrium (LTE), a poor approximation at high effective temperatures \citep{napiwotzki1997}. 

In light of this situation, the primary aim of the present paper is to improve our understanding of the spectral evolution of hot white dwarfs by performing an exhaustive, homogeneous, state-of-the-art model-atmosphere analysis of a large sample of stars drawn from the SDSS. Ultimately, we seek to answer a number of questions that remain unsettled to this day: What fraction of all white dwarfs are born with a helium atmosphere? Among those, how many eventually develop a hydrogen-rich atmosphere, and how many retain a helium-rich atmosphere throughout their life? How does the number of hybrid white dwarfs, both with homogeneous and stratified atmospheres, vary with effective temperature? What does this imply about the role of stellar winds and the helium-to-hydrogen transition? What is the total hydrogen content of these various groups of white dwarfs, and how will it impact their future spectral evolution?

The paper is organized as follows. In Section \ref{sec:sample}, we describe our sample selection as well as the observational data used in our analysis. In Section \ref{sec:theory}, we outline our theoretical framework consisting of new white dwarf model atmospheres and evolutionary sequences. Section \ref{sec:analysis} is dedicated to our spectroscopic analysis, including the description of the fitting method and the presentation of representative fits. In Section \ref{sec:results}, we report our results regarding the physical properties of our sample and discuss their implications for the theory of the spectral evolution of white dwarfs. Finally, our conclusions are summarized in Section \ref{sec:conclu}.

\section{Sample} \label{sec:sample}

We constructed our sample of hot white dwarfs (which, we remind the reader, are defined here as those with $\Teff \ge 30,000$ K) starting from the available SDSS white dwarf catalogs, up to the DR12\footnote{Note that the most recent data release of the SDSS is DR16 \citep{ahumada2020}.} \citep{kleinman2013,kepler2015,kepler2016}. In a first step, we selected only the bluest objects by applying the following simple color criteria to the $ugriz$ photometry given in the catalogs: $u-g < 0$, $u < 21$, and $g < 21$. These restrictions resulted in a raw sample of 6270 stars. In a second step, we retrieved the optical spectra of these objects from the SDSS database and performed preliminary fits to all spectra with both our pure-hydrogen and pure-helium model atmospheres (see Sections \ref{sec:theory} and \ref{sec:analysis}) in order to clean up our initial sample. Several stars turned out to be hot subdwarfs with $\logg < 6.5$ or white dwarfs with $\Teff < 30,000$ K, and were thus rejected. We assigned spectral types to the remaining objects based on a careful visual inspection of each spectrum. Our final sample contains 1806 white dwarfs, including 1638 DA, 95 DO, and 73 DB stars\footnote{In the canonical classification scheme, helium-atmosphere white dwarfs are assigned the spectral type DB if they show only He {\sc i} lines or the spectral type DO if they show at least one He {\sc ii} line, whatever its strength \citep{sion1983,wesemael1993}. In the present paper, we adopt a slightly different convention. If both He {\sc i} and He {\sc ii} features are present in a spectrum, our spectral designation is based on the strongest set of lines: we use the letter B if the He {\sc i} lines dominate or the letter O if the He {\sc ii} lines dominate. Note that a similar approach was used by \citet{krzesinski2004} and \citet{manseau2016}.}. Among these, we identified 127 hybrid white dwarfs (whose spectra show both hydrogen and helium lines), namely, 96 DAO, 6 DAB, 8 DOA, and 17 DBA stars. In addition, 29 of our DO white dwarfs exhibit traces of metals through a weak C {\sc iv} $\lambda$4658 feature, and are therefore members of the DOZ spectral class. Finally, 185 objects in our sample have M dwarf companions that contaminate the red portion of their optical spectra.

\begin{figure}
\centering
\includegraphics[width=0.975\columnwidth,clip=true,trim=2.3cm 4.9cm 1.8cm 5.3cm]{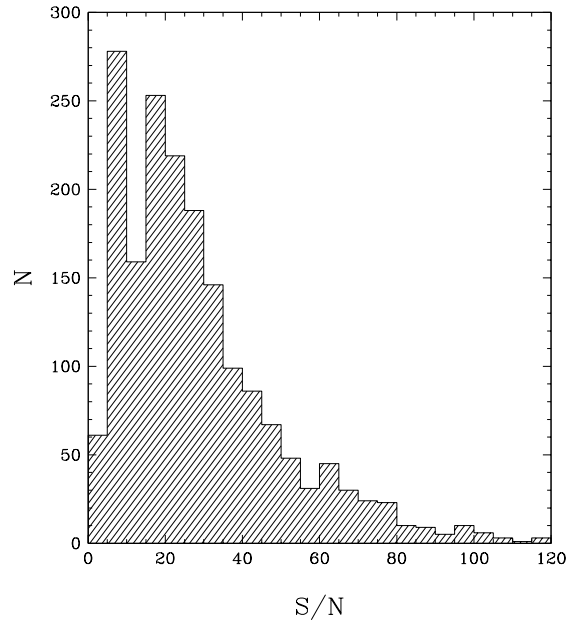}
\caption{Distribution of spectral signal-to-noise ratios for our sample of 1806 white dwarfs.}
\vspace{2mm}
\label{fig:N_SNR}
\end{figure}

Figure \ref{fig:N_SNR} shows the distribution of signal-to-noise ratios (S/N) of the spectroscopic observations, computed in the featureless range 5100 \AA\ $< \lambda <$ 5300 \AA. As expected, our sample is dominated by faint objects and hence low-S/N data. Indeed, in the SDSS, the brightness and S/N are correlated, because the same exposure time is set for all targets on a given plate. We note that 1467 of our 1806 white dwarf spectra have S/N $\ge$ 10.

In addition to spectroscopy, we also gathered $ugriz$ photometry from the SDSS database for all stars in our sample with the aim of studying their spectral energy distribution. We chose not to rely further on the $ugriz$ magnitudes given in the white dwarf catalogs mentioned above, because the SDSS photometric calibration has been updated since their publication. Furthermore, we cross-matched our sample with the {\it Gaia} DR2 catalog \citep{gaia2018} and could retrieve trigonometric parallax measurements for 1721 objects, including 824 with $\sigma_\pi/\pi \le 25\%$. In the remainder of the paper, when we make use of the astrometric and photometric observations, we confine our attention to the latter subset with high-quality data. We obtained distance estimates by simply inverting the parallaxes, as this procedure agrees well with more sophisticated probabilistic methods for small parallax uncertainties \citep{bailer-jones2015,bailer-jones2018}. Using the extinction maps of \citet{schlafly2011} together with the {\it Gaia} distances, we dereddened the $ugriz$ magnitudes following the approach outlined in \citet{harris2006}. We also applied the SDSS-to-AB system corrections to the $u$, $i$, and $z$ bands suggested by \citet{eisenstein2006b}.

\begin{figure}
\centering
\includegraphics[width=0.975\columnwidth,clip=true,trim=2.3cm 4.9cm 1.8cm 5.3cm]{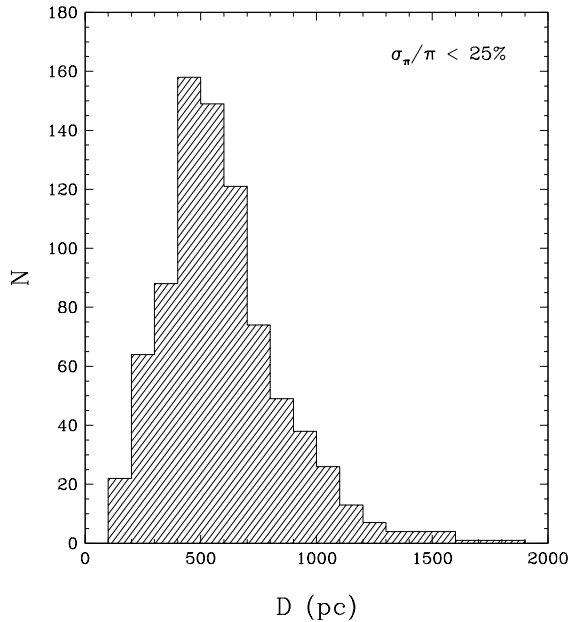}
\caption{Distribution of distances for our subsample of 824 white dwarfs with $\sigma_\pi/\pi \le 25\%$.}
\vspace{2mm}
\label{fig:N_D}
\end{figure}

Figure \ref{fig:N_D} displays the distribution of distances for our subsample of objects with precise parallax measurements. Unsurprisingly, our hot white dwarfs are on average quite distant: most of them are located between 200 and 1000 pc from the Sun (and none of them are part of the 100 pc solar neighborhood). Besides, the large-distance tail of the histogram is certainly underestimated as a result of our parallax error cut. The shape of the distribution originates from two competing factors: on one hand, hot white dwarfs are locally much rarer than their cooler counterparts because of their shorter evolutionary timescales; on the other hand, they are intrinsically more luminous and can thus be detected farther away. Figure \ref{fig:N_D} underlines the fact that interstellar reddening probably affects our entire sample.

\begin{figure*}
\centering
\includegraphics[width=1.080\columnwidth,angle=270,clip=true,trim=4.3cm 2.6cm 3.8cm 1.0cm]{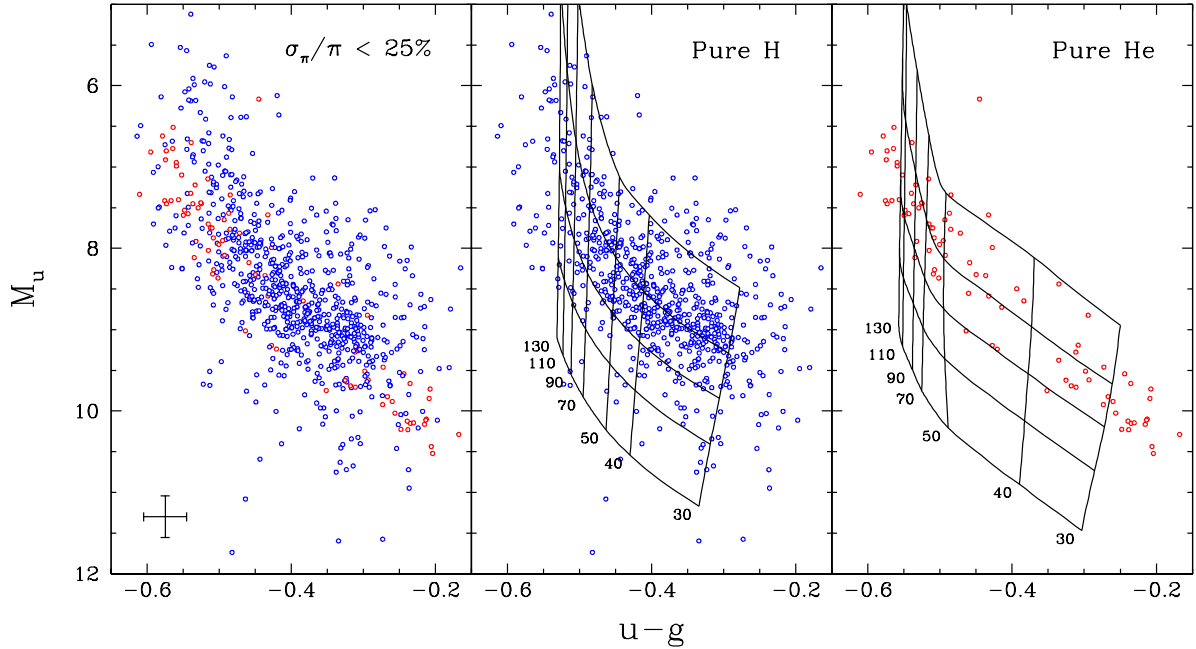}
\caption{Dereddened $M_u$ vs. $u-g$ color-magnitude diagram for our subsample of 824 white dwarfs with $\sigma_\pi/\pi \le 25\%$. Hydrogen-rich and helium-rich objects are shown in blue and red, respectively. The error bars represent the average uncertainties. The middle and right panel display sequences of theoretical colors for pure-hydrogen and pure-helium atmospheres, respectively, with masses of 0.4, 0.6, 0.8, 1.0, and 1.2 \msun\ (from top to bottom) and effective temperatures indicated in units of $10^3$ K.} 
\vspace{2mm}
\label{fig:Mu_u-g}
\end{figure*}

The dereddened $M_u$ vs. $u-g$ color-magnitude diagram for the same subsample is presented in Figure \ref{fig:Mu_u-g}, where hydrogen-rich and helium-rich white dwarfs are shown in blue and red, respectively. Also displayed are sequences of theoretical colors for both pure-hydrogen and pure-helium atmospheres, computed using our new model atmospheres and cooling tracks introduced in Section \ref{sec:theory} below. Most hydrogen-rich and helium-rich objects cluster around the corresponding 0.6 \msun\ sequences, as expected. However, there is a clear distinction between the two rightmost panels of Figure \ref{fig:Mu_u-g}: the DA population contains a significant number of overluminous stars, which are either very low-mass white dwarfs or unresolved double degenerate systems, while very few such objects are found in the DO/DB population. This dichotomy is also observed in cooler white dwarf samples \citep{genest-beaulieu2019a,bergeron2019} and probably means that evolutionary processes producing low-mass white dwarfs or double degenerate binaries almost always lead to the formation of hydrogen atmospheres.

Since hot stars emit most of their light in the ultraviolet, optical photometry samples the Rayleigh-Jeans tail of their spectral energy distribution and is therefore extremely weakly sensitive to their effective temperature. This is highlighted by the theoretical sequences illustrated in Figure \ref{fig:Mu_u-g}: the synthetic $u-g$ color index varies very slowly with temperature (for a 0.6 \msun\ hydrogen-atmosphere white dwarf, the difference in $u-g$ between $\Teff =$ 50,000 and 100,000 K is a mere 0.06 mag) and even tends asymptotically to a minimum value at the hot end. Consequently, the $ugriz$ magnitudes cannot be employed to measure the effective temperatures of the objects in our sample. Nonetheless, when coupled with parallax measurements, they can still provide information on the stellar masses: for a given temperature, the synthetic $M_u$ absolute magnitude strongly depends on the mass, as shown in the color-magnitude diagram. In other words, for our hot white dwarfs, the combination of SDSS photometry and {\it Gaia} astrometry cannot be used to determine the effective temperatures, but can be used to estimate the masses (or, alternatively, the surface gravities) if the temperatures are known from another source, such as spectroscopy.

A curious aspect of Figure \ref{fig:Mu_u-g} is that some stars have a $u-g$ color index redder than that corresponding to our low-temperature selection limit of $\Teff = 30,000$ K, or bluer than the high-temperature asymptotic value. In the former case, the main cause is the well-established disparity between the spectroscopic and photometric temperature scales: while all selected white dwarfs have $\Teff \ge 30,000$ K based on spectroscopy, some of them may have $\Teff < 30,000$ K based on photometry, as photometric temperatures are on average lower than spectroscopic temperatures by 5$-$10\% \citep{tremblay2019b,genest-beaulieu2019a,bergeron2019}. In addition, a few white dwarfs look redder than they really are because their $g$ magnitudes are contaminated by main-sequence companions. In the latter case, the reason why the bluest objects appear too blue is most likely related to the SDSS-to-AB system correction of $\Delta u = -0.04$ mag taken from \citet{eisenstein2006b}. While this correction is adequate for the majority of the stars in our sample, Figure \ref{fig:Mu_u-g} indicates that it is too large for the hottest white dwarfs, an outcome that is not too surprising since the offset between the SDSS and AB magnitude systems was calibrated using cooler objects \citep{eisenstein2006b}.

\newpage

\section{Theoretical Framework} \label{sec:theory}

\subsection{Model Atmospheres} \label{sec:theory_atmo}

We employed the publicly available codes TLUSTY and SYNSPEC \citep{hubeny1995,hubeny2017a,hubeny2017b,hubeny2017c} to compute non-LTE model atmospheres and synthetic spectra of hot white dwarfs. We considered one-dimensional, plane-parallel structures in hydrostatic, statistical, and radiative equilibrium. This last assumption, namely, the neglect of convective energy transport, is appropriate in the present context, with a small exception discussed below. 

Starting from version 205 of TLUSTY and version 51 of SYNSPEC, we made two major improvements to the codes regarding the treatment of the He {\sc i} opacity. First, we incorporated the state-of-the-art Stark profiles of \citet{beauchamp1997} for He {\sc i} lines. These profiles, which supersede the best option offered until now (a combination of data from \citealt{shamey1969}, \citealt{barnard1974} and \citealt{dimitrijevic1984}), are essential to properly model the broad absorption features of helium-rich white dwarfs. Second, we refined the implementation of the occupation probability formalism of \citet{hummer1988} and the corresponding pseudo-continuum opacity by dropping the assumption of a hydrogenic ion (see \citealt{beauchamp1995} for details). Regarding other important ionic species, we used the detailed Stark profiles of \citet{tremblay2009} and \citet{schoening1989} for H {\sc i} and He {\sc ii} lines, respectively, both already included in TLUSTY and SYNSPEC. Note that we applied the Hummer-Mihalas formalism to these two ions as well.

\begin{figure}
\centering
\includegraphics[width=0.975\columnwidth,clip=true,trim=1.2cm 6.2cm 1.0cm 6.0cm]{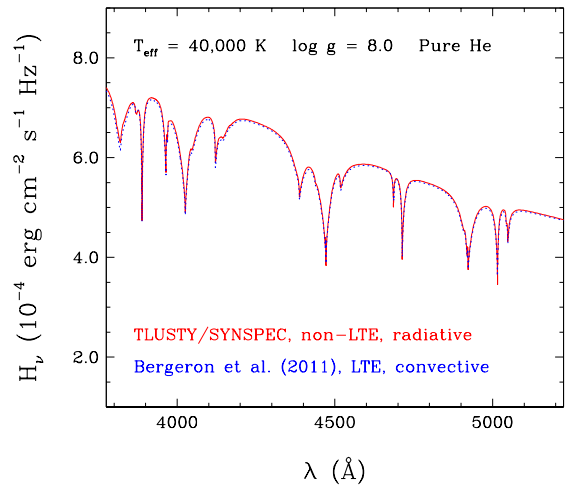}
\caption{Comparison of synthetic optical spectra generated from two different white dwarf model atmospheres with $\Teff = 40,000$ K, $\logg = 8.0$, and a pure-helium composition: a non-LTE, strictly radiative model computed with the codes TLUSTY/SYNSPEC (solid red line), and a LTE, convective model computed with the code of \citealt{bergeron2011} (dotted blue line).}
\vspace{2mm}
\label{fig:synth_comp}
\end{figure}

We calculated a pure-hydrogen grid and a pure-helium grid, which both cover the following parameter space: $30,000 \ {\rm K} \le \Teff \le 150,000$ K (in steps of 2500 K for $30,000 - 60,000$ K, 5000 K for $60,000 - 90,000$ K, and 10,000 K for $90,000 - 150,000$ K) and $6.5 \le \logg \le 9.5$ (in steps of 0.5 dex). In the pure-hydrogen models, the neglect of convection is an excellent approximation over the entire temperature range of interest here, since the atmospheres of DA stars become convective only for $\Teff \ \lta 15,000$ K (see Figure 5 of \citealt{tremblay2010}). However, this is not true for the pure-helium models: a thin convection zone develops in the atmospheres of DO/DB stars for $\Teff \ \lta 60,000$ K due to the partial ionization of He {\sc ii} (see Figure 3 of \citealt{bergeron2011}). Unfortunately, strong numerical instabilities arise when convective energy transport and non-LTE effects are included simultaneously in helium-rich TLUSTY models. Nevertheless, the LTE calculations performed by \citet{bergeron2011} indicate that convection becomes truly significant only close to the low-temperature limit of our grid, where non-LTE effects are unimportant. For instance, in their pure-helium models with $\logg = 8.0$, the convective flux accounts for at most 0.02\%, 6\%, and 57\% of the total flux at $\Teff =$ 45,000, 40,000, and 35,000 K, respectively. Consequently, for our pure-helium grid, we adopted the following strategy: we employ the non-LTE, strictly radiative models of TLUSTY/SYNSPEC for $\Teff > 40,000$ K, and the LTE, convective models of \citet{bergeron2011} for $\Teff \le 40,000$ K. We verified that the two sets of synthetic spectra are in good agreement at the branching point for every $\logg$ value, as illustrated in Figure \ref{fig:synth_comp} for $\logg = 8.0$.

For the purpose of analyzing our hybrid white dwarfs, we also calculated model atmospheres containing both hydrogen and helium. We constructed two separate grids differing by the assumed chemical configuration, which we refer to as homogeneous and stratified. In our chemically homogeneous models, hydrogen and helium are mixed uniformly in the atmosphere, in which case the helium-to-hydrogen number ratio (He/H) is a free parameter. Our homogeneous grid spans the same effective temperatures and surface gravities as our pure-composition grids, as well as abundances of $-5.0 \le \log {\rm He/H} \le 5.0$ (in steps of 1.0 dex). For helium-rich ($\log {\rm He/H} > 0$) models, we adopted the same compromise between non-LTE effects and convection as above.

In our chemically stratified models, a thin hydrogen layer floats on top of a helium envelope in diffusive equilibrium, in which case the mass of the superficial hydrogen layer ($q_{\rm H} \equiv M_{\rm H}/M_{\star}$) is a free parameter. The equilibrium chemical profile was obtained from the formalism of \citet{vennes1988}, as implemented in TLUSTY and SYNSPEC by \citet{barstow1998a}. To our knowledge, this is the first time that an exhaustive set of such non-LTE stratified model atmospheres is presented. The parameter space covered by our stratified grid is $30,000 \ {\rm K} \le \Teff \le 90,000$ K (in steps of 2500 K for $30,000 - 60,000$ K, and 5000 K for $60,000 - 90,000$ K), $7.0 \le \logg \le 9.0$ (in steps of 0.5 dex), and $-13.0 \le Q \le -5.0$ (in steps of 1.0) where $Q \equiv 3 \logg + 2 \log q_{\rm H}$. The introduction of the quantity $Q$ allows us to consider a different range of $\log q_{\rm H}$ values for each $\logg$ value, and its definition is such that synthetic spectra with a given $Q$ value look fairly similar to each other \citep{manseau2016}. For example, at $\logg = 8.0$, hydrogen layer masses are comprised in the range $-18.5 \le \log q_{\rm H} \le -14.5$. Note that we could not follow our usual approach for taking convection into account, since no convective stratified models are available in the literature. Hence, one must keep in mind that the cool ($\Teff \ \lta 40,000$ K) models of our stratified grid are approximate (see \citealt{manseau2016} for a discussion of the occurrence of convection in chemically layered atmospheres). 

\begin{figure*}
\centering
\includegraphics[width=1.875\columnwidth,clip=true,trim=2.0cm 6.1cm 1.4cm 5.2cm]{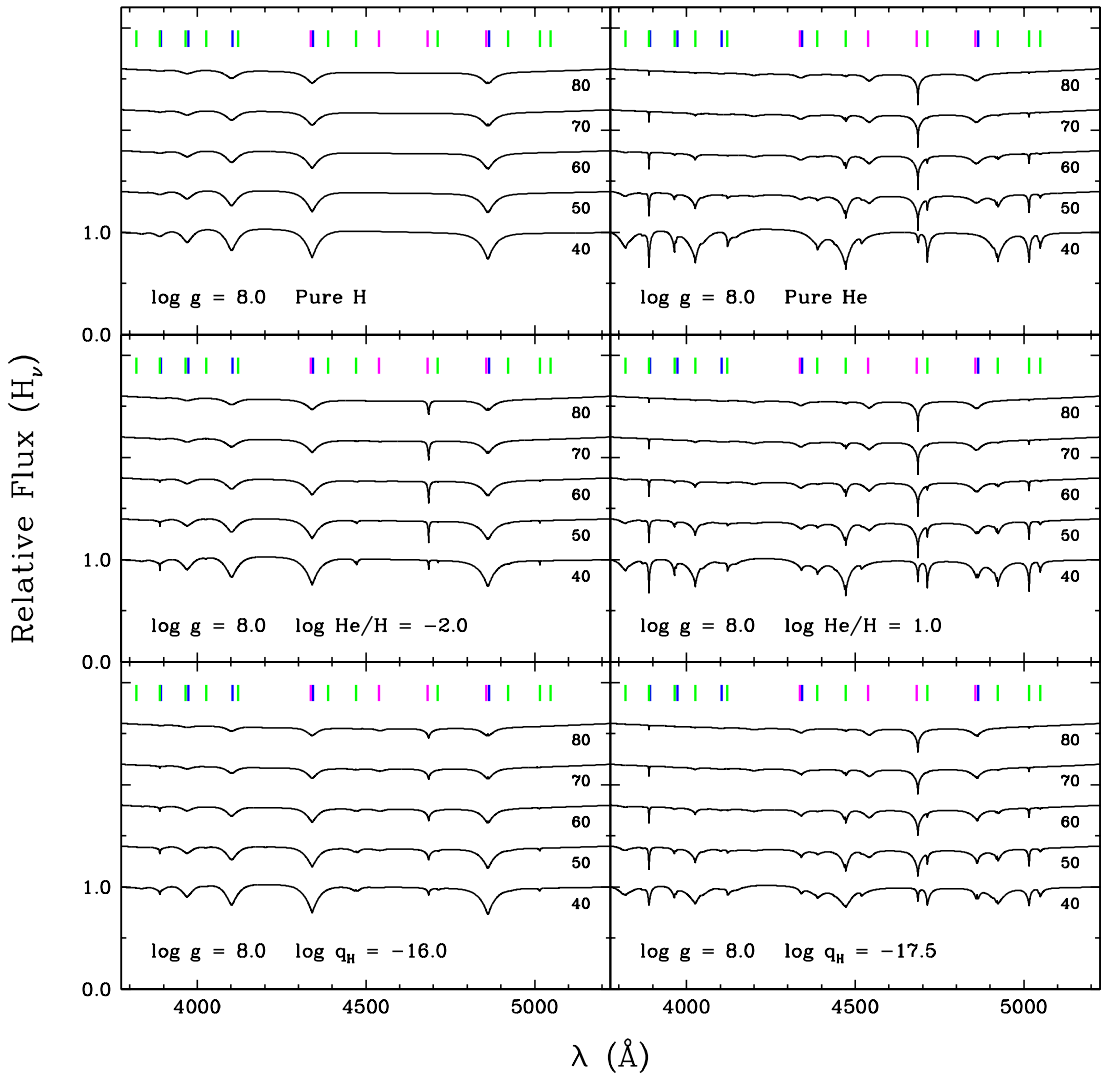}
\caption{Synthetic optical spectra of white dwarf model atmospheres at $\logg = 8.0$, for various effective temperatures and chemical compositions. The temperatures are indicated in units of $10^3$ K, and the compositions are: pure hydrogen (top left panel), pure helium (top right panel), hydrogen and helium in a homogeneous configuration (middle panels), and hydrogen and helium in a stratified configuration (bottom panels). The spectra are normalized to a continuum set to unity and offset vertically from each other by 0.4 for clarity. The positions of the H {\sc i}, He {\sc i}, and He {\sc ii} lines are marked by blue, green, and magenta ticks, respectively.}
\vspace{2mm}
\label{fig:synth}
\end{figure*}

Finally, we relied on the non-LTE model atmospheres and synthetic spectra introduced by \citet{gianninas2010} to analyze the objects afflicted by the Balmer-line problem. These models, also computed with TLUSTY and SYNSPEC, contain carbon, nitrogen, and oxygen at solar abundances (in a homogeneous configuration) and incorporate the appropriate Stark broadening for these elements, as prescribed by \citet{werner1996a}. While this addition effectively solves the Balmer-line problem, we stress that such a large amount of CNO is not observed in hot white dwarfs and simply acts as a proxy for all metals here. It should be mentioned that our own calculations and those of \citet{gianninas2010} made use of the same input physics, with the obvious exception of our improvements to the TLUSTY and SYNSPEC codes described above. However, these modifications only affect the He {\sc i} opacity, which is negligible in the hot, hydrogen-rich stars suffering from the Balmer-line problem.

Figure \ref{fig:synth} presents some of our synthetic spectra at $\logg = 8.0$, for selected effective temperatures and atmospheric compositions. The top left and top right panels show spectra generated from, respectively, pure-hydrogen and pure-helium models (typical of DA and DO/DB white dwarfs). As the temperature decreases, the spectral appearance reflects the changing ionization balance: the H {\sc i} lines strengthen continuously; the He {\sc ii} lines first strengthen, reach a maximum intensity around $\Teff \sim 65,000$ K, and then weaken; the He {\sc i} lines appear around $\Teff \sim 75,000$ K and then strengthen steadily. The middle and bottom panels display spectra corresponding to, respectively, homogeneous and stratified models. In each instance, the left panel shows spectra dominated by hydrogen features (typical of DAO/DAB stars), whereas the right panel shows spectra exhibiting primarily helium features (typical of DOA/DBA stars). The chemical structure of the atmosphere clearly influences the shape of the absorption lines, which are much broader and shallower in the stratified case than in the homogeneous case (see \citealt{manseau2016} for more on this topic). Finally, the right panels of Figure \ref{fig:synth} highlight the well-known difficulty of detecting a small amount of hydrogen in a hot helium-rich atmosphere, every H {\sc i} feature being blended with a He {\sc ii} feature \citep{werner1996b}.

\subsection{Evolutionary Cooling Sequences} \label{sec:theory_seq}

In order to convert atmospheric parameters (effective temperature and surface gravity) into stellar properties (mass, radius, luminosity, and cooling age), one needs a set of detailed white dwarf evolutionary sequences. The Montreal sequences, first introduced by \citet{fontaine2001} and available online at \url{http://www.astro.umontreal.ca/~bergeron/CoolingModels/} (in the section titled ``Evolutionary Sequences''), are widely employed in the literature for that purpose. However, in their latest version, they are not suitable for the present work, because their validity is restricted to relatively cool ($\Teff \ \lta 30,000$ K) degenerates. The main reason is that they were computed starting from static white dwarf models, which represent a poor approximation at high effective temperatures, when gravitational contraction and neutrino emission are still significant. Therefore, we decided to calculate a whole new set of white dwarf cooling tracks using more realistic models as initial conditions and thus extending to much higher effective temperatures.

To generate our starting models, we relied on the public stellar-evolution code Modules for Experiments in Stellar Astrophysics (MESA; \citealt{paxton2011,paxton2013,paxton2015,paxton2018,paxton2019}). We evolved 14 stellar models with different initial masses from the zero-age main sequence (ZAMS) to the beginning of the white dwarf stage. We considered initial masses ranging from 0.4 to 11 \msun\ (corresponding to final masses ranging from 0.306 to 1.223 \msun) and an initial metallicity of 0.02 in all cases. We adopted the mass-loss prescriptions of \citet{reimers1975} on the red giant branch (RGB) and \citet{blocker1995} on the asymptotic giant branch (AGB), both with an efficiency parameter ($\eta$) of 0.5, except for stars with high initial masses ($\ge$4 \msun), for which we increased the AGB efficiency parameter to 20. As pointed out in previous studies utilizing MESA, this artificial enhancement of mass loss is necessary to avoid numerical difficulties associated with the occurrence of thermal pulses on the AGB \citep{paxton2013,lauffer2018}. Other computational settings were chosen similar to those of the test-suite case {\tt 1M\_pre\_ms\_to\_wd}.

Next, to calculate state-of-the-art white dwarf cooling sequences, we turned to our own evolutionary code, designed specifically for the modeling of white dwarfs. We used a revamped, modern implementation of our code named STELUM, for STELlar modeling from Universit\'e de Montr\'eal. This program offers a variety of applications geared mostly toward chemical evolution and asteroseismology (see \citealt{brassard2018} for an early report). An elaborate description of STELUM is clearly beyond the scope of this work and is thus postponed to a future publication (B\'edard et al.~2020, in preparation). It is nonetheless worth mentioning that STELUM produces and evolves complete models of stars, down from the center up to the very surface. The constitutive physics of our code is broadly similar to that outlined in \citet{fontaine2001} and \citet{vangrootel2013}, apart from the treatment of the conductive opacities, which are now taken from \citet{cassisi2007} instead of from \citet{hubbard1969} and \citet{itoh1983,itoh1984,itoh1993}. The effects of this improvement on the cooling process are discussed below.

We chose to follow the same approach as in \citet{fontaine2001}, that is, to compute evolutionary tracks with pre-specified masses and chemical compositions (as opposed to those obtained from the MESA calculations). We assumed a core made of a uniform mixture of carbon and oxygen in equal proportions ($X_{\rm C} = X_{\rm O} = 0.5$), surrounded by a helium mantle and an outermost hydrogen layer. We considered two standard envelope compositions differing by the thickness of the hydrogen layer, which we refer to as thick ($q_{\rm He} = 10^{-2}$, $q_{\rm H} = 10^{-4}$) and thin ($q_{\rm He} = 10^{-2}$, $q_{\rm H} = 10^{-10}$). For each of these two chemical profiles, we considered 23 mass values covering the range 0.2 \msun\ $\le M \le$ 1.3 \msun\ (in steps of 0.05 \msun), for a total of 46 cooling tracks. Of course, low-mass ($M \ \lta 0.45$ \msun) and high-mass ($M \ \gta 1.1$ \msun) degenerates are expected to harbor helium cores and oxygen/neon cores, respectively, instead of carbon/oxygen cores; for these masses, our computations should hence be viewed with caution. To produce a given sequence, we proceeded as follows. We first built an approximate initial model by combining a pre-white dwarf thermodynamic structure obtained from MESA (interpolated at the desired mass) with our fixed chemical structure. This model was then fed into STELUM and relaxed for a few iterations, to allow the thermodynamic structure to adjust to the new chemical structure. Finally, this self-consistent model was used as input for the calculation of the detailed evolutionary sequence. In order to keep the composition constant with cooling, diffusive equilibrium was assumed (that is, transport mechanisms were turned off) and residual nuclear burning was neglected. While the strategy of forcing a specific chemical profile might seem somewhat arbitrary, it does have the advantage of minimizing the dependence of our results on the numerous uncertainties of pre-white dwarf evolution, most notably regarding core convective overshooting, helium-burning reaction rates, and wind mass-loss rates \citep{salaris2010,degeronimo2017}.

\begin{figure}
\centering
\includegraphics[width=0.975\columnwidth,clip=true,trim=2.3cm 4.9cm 1.8cm 3.3cm]{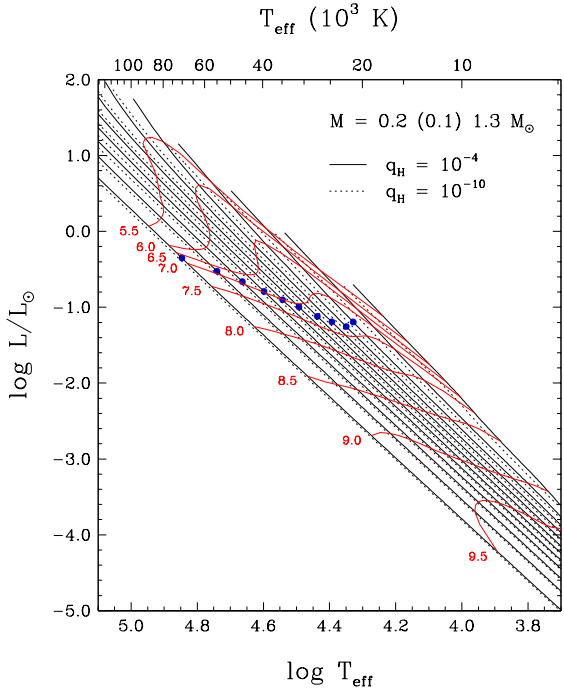}
\caption{Evolutionary sequences of white dwarf models in the theoretical Hertzsprung-Russell diagram, for masses of 0.2 to 1.3 \msun\ in steps of 0.1 \msun\ (from top to bottom), and for thick (solid black lines) and thin (dotted black lines) hydrogen layers. Also shown, for the thick-layer models only, are isochrones labeled in units of $\log \tau_{\rm cool}$, where $\tau_{\rm cool}$ is the white dwarf cooling age in yr (solid red lines), as well as the location of the transition between the neutrino and thermal cooling phases (blue dots).}
\vspace{2mm}
\label{fig:seq}
\end{figure}

Figure \ref{fig:seq} shows some of our new evolutionary tracks, for various masses and for both envelope compositions, in the theoretical Hertzsprung-Russell (HR) diagram, together with representative isochrones for the thick-layer models only. It is readily seen that our sequences now extend to much higher effective temperatures than before, most of them beginning at $\Teff > 100,000$ K (except, of course, for the low-mass models, which never reach such high temperatures in their evolution). The cooling of a young, hot white dwarf is essentially driven by the emission of neutrinos formed in large numbers in the hot stellar core. These weakly interacting particles escape directly from the central regions to the outer space, thus providing an extremely efficient energy-loss mechanism and making this early evolution quite rapid. As cooling proceeds, the neutrino luminosity decreases and eventually drops below the photon luminosity, at which point the evolution becomes dominated by the much slower process of thermal cooling. The location of this transition on the evolutionary tracks is shown by the blue dots in Figure \ref{fig:seq}, again for the thick-layer models only. (Note that no blue dots are displayed on the very low-mass sequences, because the photon luminosity always exceeds the neutrino luminosity in these models.) As a white dwarf leaves the neutrino cooling phase and enters the thermal cooling phase, its evolutionary rate slows down, as illustrated by the tightening of the isochrones near the blue dots in Figure \ref{fig:seq}. The effective temperature of the transition sensitively depends on the stellar mass: for more massive stars, the neutrino-to-photon luminosity ratio declines at higher temperatures, which explains the shape of the isochrones in the theoretical HR diagram.

\begin{figure}
\centering
\includegraphics[width=0.975\columnwidth,clip=true,trim=3.1cm 2.5cm 2.6cm 2.5cm]{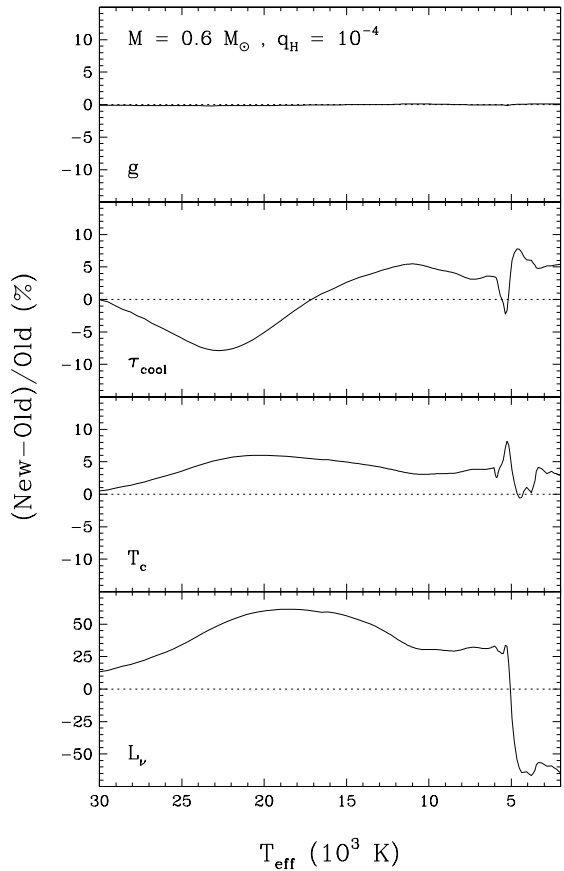}
\caption{Percentage difference in surface gravity, cooling age, central temperature, and neutrino luminosity (from top to bottom) as a function of effective temperature, between our old and new 0.6 \msun, thick-layer evolutionary sequences, which differ by the treatment of the conductive opacity.}
\vspace{2mm}
\label{fig:seq_comp}
\end{figure}

For completeness, we now make a brief digression from our main subject to discuss differences between our old and new evolutionary sequences at low effective temperatures ($\Teff < 30,000$ K). To this end, as an illustrative example, we display in Figure \ref{fig:seq_comp} the percentage change of various physical quantities (namely, the surface gravity, the cooling age, the central temperature, and the neutrino luminosity) as a function of effective temperature, for the typical case $M = 0.6$ \msun, $q_{\rm H} = 10^{-4}$. The primary source of discrepancy between our previous and current generations of calculations at low temperatures is the inclusion of the conductive opacities of \citet{cassisi2007} in STELUM. On one hand, this improvement has absolutely no effect on the mechanical properties of white dwarf models. In fact, for all cooling tracks, the changes in surface gravity, radius, and luminosity are smaller than 0.5\% over the entire temperature range of validity of our old models. This is illustrated in the first panel of Figure \ref{fig:seq_comp} for the particular case of the surface gravity in our reference sequence. Therefore, the theoretical mass-radius relation is basically unaltered, and the conclusions achieved by recent studies on its accuracy remain intact \citep{bedard2017,parsons2017,tremblay2019b,genest-beaulieu2019a}. 

On the other hand, the new conductive opacities substantially impact the thermal properties of white dwarf models. As shown by \citet{salaris2013}, the opacities of \citet{cassisi2007} are larger than those of \citet{hubbard1969} and \citet{itoh1983,itoh1984,itoh1993}, both in the carbon/oxygen core and the helium envelope. One could naively expect the enhanced opacity to simply slow down the cooling, as a result of the reduced efficiency of radiative transfer throughout the star. In other words, it would be reasonable to expect the cooling time required to reach a given effective temperature to be longer. This hypothesis is proven wrong by the second panel of Figure \ref{fig:seq_comp}: the cooling age is actually shorter at high temperatures and longer at low temperatures, by as much as $\sim$10\%. The correct explanation for this behavior was first put forward by \citet{salaris2013}. The updated conductive opacities cause the core temperature to rise by a few percents, as displayed in the third panel of Figure \ref{fig:seq_comp}. Since neutrino production rates strongly depend on temperature, this in turn leads to an increase in neutrino luminosity as large as $\sim$50\%, as shown in the fourth panel of Figure \ref{fig:seq_comp}. Consequently, at high effective temperatures, when neutrino emission is still the main energy sink, the cooling process is actually faster. However, as the temperature decreases and thermal cooling becomes dominant, the enhanced opacity translates into a slower evolution, as anticipated from the simple argument made above. 

Finally, it is worth mentioning that these changes in the thermal properties of our models obviously have repercussions on the crystallization process. More specifically, for most masses, the net effect of the improved conductive opacities is to shift the onset of core crystallization to slightly lower effective temperatures. This difference between our old and new sequences manifests itself as small spikes near $\Teff \sim 5500$ K in the two middle panels of Figure \ref{fig:seq_comp}. Nevertheless, we want to emphasize that these shifts are small (typically a few hundred K) and hence do not affect the conclusions drawn by \citet{tremblay2019a} and \citet{bergeron2019} about the so-called {\it Gaia} crystallization sequence using our previous generation of evolutionary tracks.

As before, we make our new white dwarf cooling sequences available to the community online at \url{http://www.astro.umontreal.ca/~bergeron/CoolingModels/}, as well as through the Montreal White Dwarf Database \citep{dufour2017}.

\section{Spectroscopic Analysis} \label{sec:analysis}

\subsection{Standard Fitting Procedure} \label{sec:analysis_normal}

Our next task was to analyze the SDSS optical spectra with our model atmospheres in order to derive the effective temperature, surface gravity, and atmospheric composition of each star in our sample. For objects showing spectral lines from only one element, we simply assumed a pure composition (that is, we applied our pure-hydrogen and pure-helium models to the analysis of DA and DO/DB white dwarfs, respectively), leaving only $\Teff$ and $\logg$ to be determined. Note that we also included in this category the few DOZ stars, for which we employed our pure-helium models as well, since we did not calculate metal-rich models, admittedly more adequate here but much more computationally expensive. The study of hybrid white dwarfs, whose chemical configuration is a priori unknown, required a more elaborate approach, the description of which is deferred to the next subsection.

We relied on a two-step fitting method wherein the observed and synthetic spectra were first normalized to a continuum set to unity and then compared to each other, so that the atmospheric parameters were constrained solely from the information contained in the line profiles \citep{bergeron1992,liebert2005,bergeron2011}. The most critical aspect of the normalization process was to properly define the continuum of the observed spectrum. To do so, we fitted the observed spectrum with the grid of theoretical spectra, convolved with a Gaussian instrumental profile, and multiplied by a sixth-order polynomial in wavelength designed to account for interstellar reddening and flux calibration errors. The optimization was carried out using the non-linear least-square Levenberg-Marquardt algorithm. Although the resulting fit did not yield reliable values of $\Teff$ and $\logg$ due to the large number of free parameters, it provided a smooth fitting function defining the continuum of the observed spectrum. Normal points were fixed at the value of this function at a few chosen wavelengths and used to normalize the spectrum to a continuum set to unity. Then, we fitted this normalized spectrum with the grid of normalized synthetic spectra, still convolved with a Gaussian instrumental profile, again employing the Levenberg-Marquardt algorithm. This time, we obtained physically meaningful atmospheric parameters, since only $\Teff$ and $\logg$ were allowed to vary in the optimization procedure. Finally, each fit was visually inspected to ensure the quality of the results.

\begin{figure*}
\centering
\includegraphics[width=1.875\columnwidth,clip=true,trim=2.7cm 6.8cm 2.0cm 6.1cm]{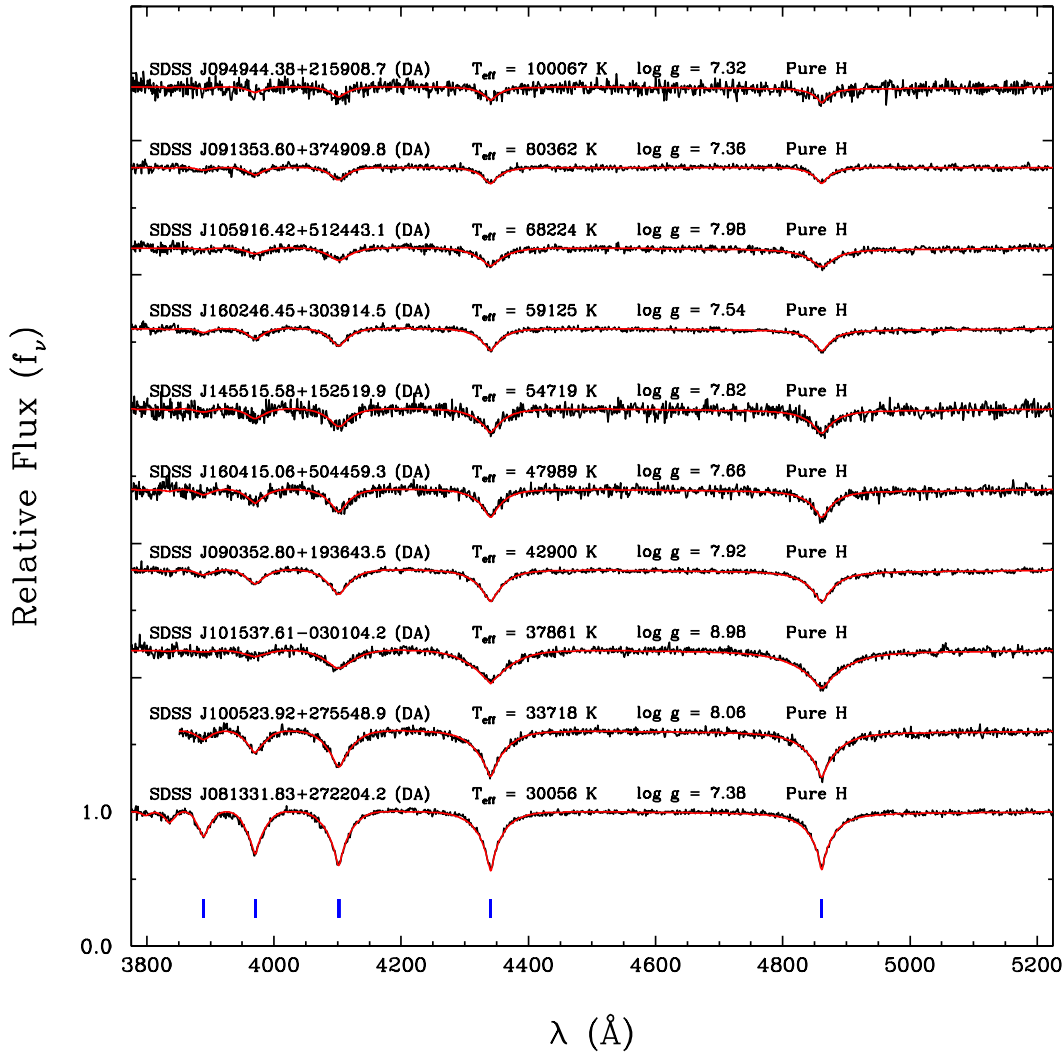}
\caption{Best model-atmosphere fits to the optical spectra of DA white dwarfs with high signal-to-noise observations (S/N $>$ 30). The observed and synthetic spectra, normalized to a continuum set to unity, are displayed as black and red lines, respectively. The fits are offset vertically from each other by 0.6 for clarity. The SDSS name, the spectral type, and the best-fitting atmospheric parameters are given for each object. The positions of the H {\sc i} lines are marked by blue ticks.}
\vspace{2mm}
\label{fig:fit_DA}
\end{figure*}

\begin{figure*}
\centering
\includegraphics[width=1.875\columnwidth,clip=true,trim=2.7cm 6.8cm 2.0cm 6.1cm]{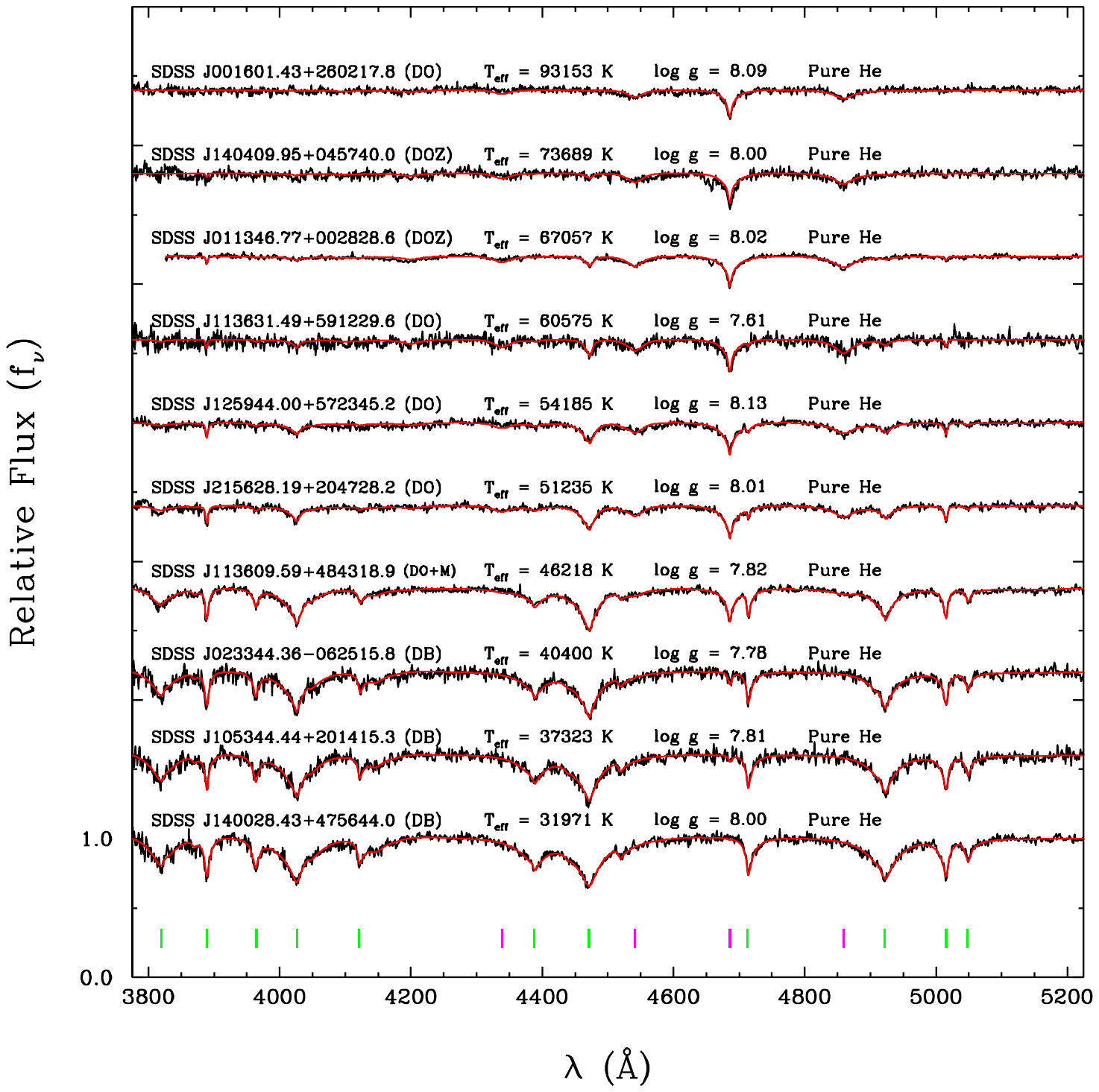}
\caption{Same as Figure \ref{fig:fit_DA}, but for DO/DB white dwarfs. The positions of the He {\sc i} and He {\sc ii} lines are marked by green and magenta ticks, respectively.}
\vspace{2mm}
\label{fig:fit_DO}
\end{figure*}

Figures \ref{fig:fit_DA} and \ref{fig:fit_DO} display our spectroscopic solutions for some DA and DO/DB white dwarfs, respectively, with high-S/N observations (S/N $>$ 30). Note that in each fit, the whole optical spectrum, up to $\lambda = 7000$ \AA, was taken into account, but we show here only the blue part for presentation purposes. In the vast majority of cases, our pure-hydrogen and pure-helium models provide excellent matches to the spectroscopic data. For the 29 DOZ stars, our pure-helium models obviously fail to reproduce the weak C {\sc iv} $\lambda$4658 feature detected in the spectra, as illustrated for two objects in Figure \ref{fig:fit_DO}. In such instances, we simply excluded this narrow wavelength region from the fits. We are confident that the neglect of the carbon opacity in our models only marginally impacts the derived values of $\Teff$ and $\logg$. Indeed, the carbon abundances measured in DOZ white dwarfs typically fall in the range $-3.0 < \log {\rm C/He} < -2.0$ \citep{dreizler1996,werner2014,reindl2014b}, and the corresponding opacity barely influences the He {\sc i} and He {\sc ii} line profiles at high temperatures according to the model-atmosphere calculations of \citet{reindl2018}.

\subsection{Hybrid White Dwarfs} \label{sec:analysis_hybrid}

Special care was needed to properly analyze the 127 stars in our sample exhibiting both hydrogen and helium spectral features. As indicated by Figure \ref{fig:synth}, the great sensitivity of the spectrum to the chemical structure of the atmosphere can be exploited to discriminate between homogeneous and stratified distributions of hydrogen and helium \citep{bergeron1994,manseau2016}. To this end, we fitted the optical spectra of our hybrid white dwarfs with both our homogeneous and stratified model atmosphere grids. The physical quantity used to measure the composition (the helium-to-hydrogen number ratio He/H in the homogeneous case, the fractional mass of the hydrogen layer $q_{\rm H}$ in the stratified case) was treated as an additional free parameter in the optimization process. We then compared the two solutions and adopted the best-fitting one.

\begin{figure*}
\centering
\includegraphics[width=1.875\columnwidth,clip=true,trim=2.7cm 6.8cm 2.0cm 6.1cm]{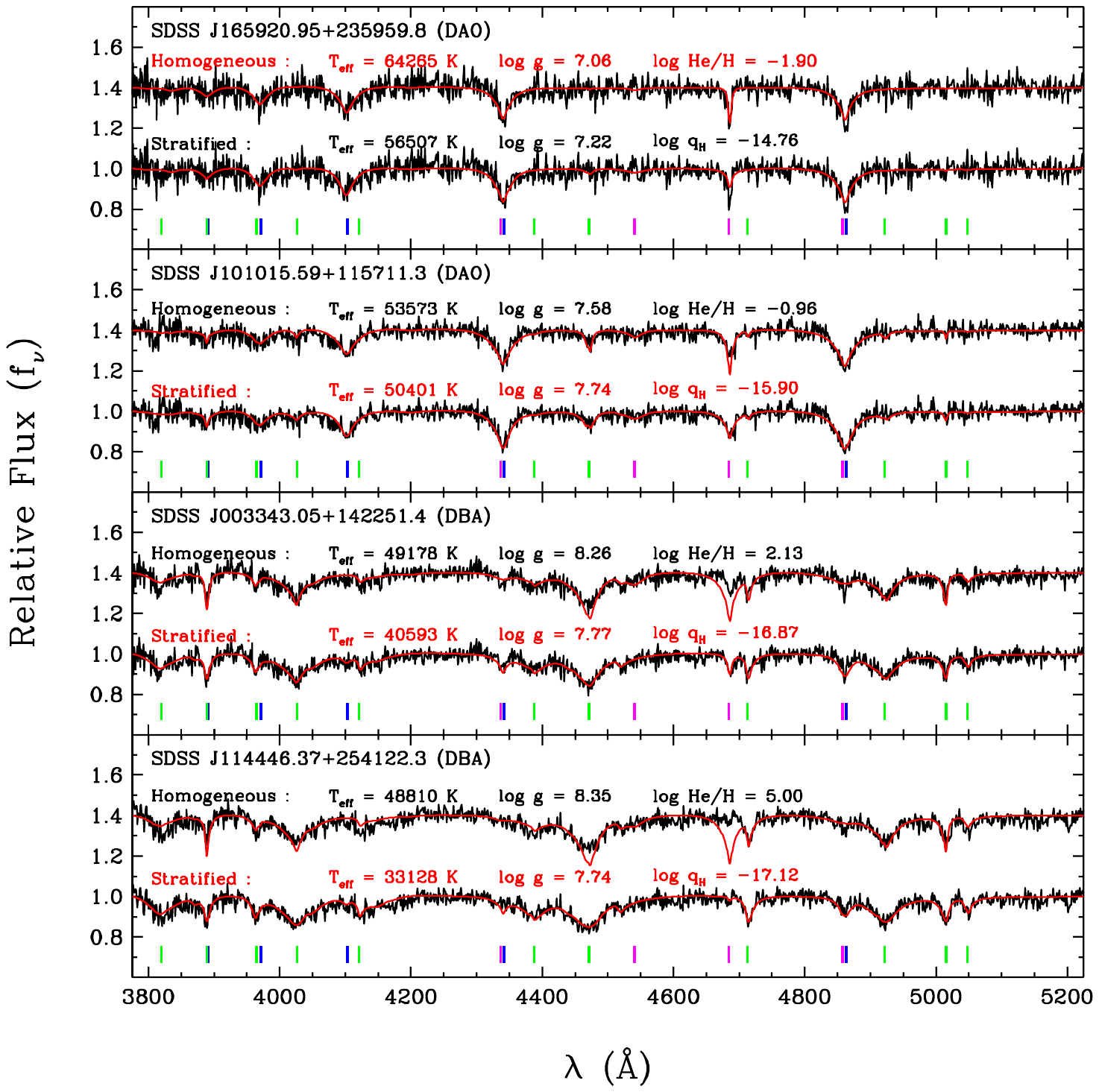}
\caption{Same as Figure \ref{fig:fit_DA}, but for hybrid white dwarfs analyzed assuming in turn a chemically homogeneous and stratified atmosphere. The adopted solution is highlighted in red. The positions of the H {\sc i}, He {\sc i}, and He {\sc ii} lines are marked by blue, green, and magenta ticks, respectively.}
\vspace{2mm}
\label{fig:fit_hyb}
\end{figure*}

Figure \ref{fig:fit_hyb} contrasts the homogeneous and stratified solutions for four hybrid white dwarfs in our sample. Once again, while only the blue optical spectra are displayed here, redder wavelengths, up to $\lambda = 7000$ \AA, were included in the fits as well. For each star, the parameters corresponding to the adopted chemical configuration are highlighted in red. The top two white dwarfs are relatively hot, in which situation the comparison rests largely on the He {\sc ii} $\lambda$4686 line, and to a lesser extent on the He {\sc i} $\lambda$4471 line. We can safely assert that the first object has a homogeneous atmosphere, since our stratified models produce a He {\sc ii} $\lambda$4686 feature that is too shallow. Conversely, the second object bears the signature of a stratified atmosphere, the He {\sc ii} $\lambda$4686 profile predicted by our homogeneous models being too deep. The bottom two panels show cooler stars, for which the distinction between both types of chemical structure is more evident. In this temperature range, most observed hydrogen and helium lines are sensitive to the chemical configuration, especially H$\alpha$, H$\beta$, and H$\gamma$, as well as He {\sc i} $\lambda$4471, $\lambda$5876, and $\lambda$6678. Moreover, when stratified white dwarfs are fitted with homogeneous models, the best solution often predicts sharp He {\sc ii} $\lambda$4686 and $\lambda$5412 features that are definitely not observed. This can be seen for the bottom two objects, which undoubtedly have stratified external layers. We want to point out that the four stars presented in Figure \ref{fig:fit_hyb} all exhibit both hydrogen and helium lines of at least moderate strength, which makes the disparity between homogeneous and stratified atmospheres quite obvious. However, this is not always the case: the difference can be more subtle when the lines of one species are much weaker than those of the other species (see Figure 13 of \citealt{manseau2016}).

Out of our 127 hybrid stars, 31 show evidence of atmospheric chemical stratification (in five cases, the evidence is somewhat ambiguous). Therefore, besides retrieving the 15 stratified white dwarfs of \citet{manseau2016}, we further identified 16 new such objects. The latter were missed by \citet{manseau2016} most likely because their search algorithm targeted SDSS white dwarfs exhibiting the He {\sc ii} $\lambda$4686 feature and consequently ignored the cooler candidates. For the stars in common, a quick comparison reveals that our effective temperatures are on average slightly lower, an effect that is probably due to our improved non-LTE model atmospheres. We note that all of our stratified white dwarfs have $\Teff < 55,000$ K, whereas most of our homogeneous white dwarfs have $\Teff > 55,000$ K. Furthermore, our estimates of the masses of hydrogen floating at the surface of stratified objects are comprised in the narrow range $-18.25 < \log q_{\rm H} < -15.25$. Of course, this result does not mean that white dwarfs with slightly lower and higher values of $q_{\rm H}$ do not exist, but rather that such stars simply show pure DO/DB and pure DA spectra, respectively, because the transition region between the hydrogen and helium layers is located far away from the photosphere. We also want to stress that the quantity $q_{\rm H}$ evaluated here cannot be interpreted as the total hydrogen content, but rather as the instantaneous amount of hydrogen floating at the surface at the present time. In fact, it is highly plausible that much more hydrogen is still diluted in the underlying helium envelope.

Finally, we mention that for two hybrid white dwarfs, SDSS J163757.58+190526.1 (DBA) and SDSS J221703.09+223330.8 (DAO), neither homogeneous nor stratified models yielded satisfactory fits to all observed spectral lines simultaneously. These two objects are actually DA+DB and DA+DO binary systems. A proper model-atmosphere study of these double degenerate binaries will be reported elsewhere.

\subsection{White Dwarfs with Main-Sequence Companions} \label{sec:analysis_M}

For some 185 white dwarfs in our sample, the optical spectra suffer from contamination by M dwarf companions, which can jeopardize the accuracy of the atmospheric parameters obtained from our fitting method. To overcome this problem, we applied the procedure developed by \citet{gianninas2011} to remove the contribution of the main-sequence stars from the observed spectra before proceeding with our standard analysis. 

Briefly, we combined our white dwarf synthetic spectra with the M dwarf spectral templates of \citet{bochanski2007}, and we fitted each observed spectrum with a function given by:
\begin{equation}
\begin{split}
F_{\rm obs} = & \left[ F_{\rm WD}(\Teff,\logg) + f \times F_{\rm M}({\rm type}) \right] \\
& \times \left[ a_0 + a_1 \lambda + a_2 \lambda^2 + a_3 \lambda^3 \right] \ . 
\nonumber
\end{split}
\end{equation}
In this equation, the flux of the white dwarf $F_{\rm WD}$ depends only on $\Teff$ and $\logg$ (for an assumed atmospheric composition), while the flux of the M dwarf $F_{\rm M}$ depends only on its spectral type, between M0 and M9. Besides these three parameters, the function also contains a scaling factor $f$ setting the relative contributions of the two stars, and the four coefficients $a_{0-4}$ of a third-order polynomial in wavelength. These eight parameters were all allowed to vary in the fitting process in order to reproduce the observed spectrum as well as possible. We took advantage of the fact that the SDSS spectroscopic data cover the near-infrared and extended the fitted wavelength range up to $\lambda = 9200$ \AA\ to put tight constraints on the M dwarf type. Once achieved, the solution provided us with the individual contributions of the white dwarf and the M dwarf to the composite spectrum. We then subtracted the synthetic flux of the main-sequence component from the observed spectrum. Finally, this decontaminated spectrum was analyzed using our standard technique outlined above to derive reliable atmospheric parameters for the white dwarf.

\begin{figure*}
\centering
\includegraphics[width=1.875\columnwidth,clip=true,trim=2.7cm 9.7cm 2.0cm 9.8cm]{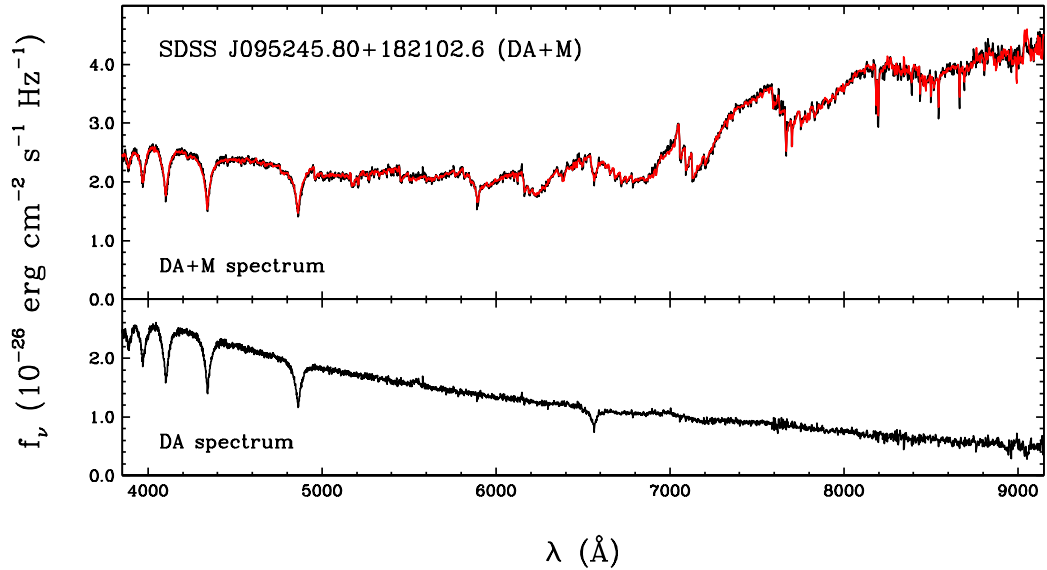}
\caption{Top panel: Best model-atmosphere fit to the optical and near-infrared spectrum of a white dwarf + M dwarf binary system. The observed and synthetic spectra are displayed as black and red lines, respectively. Bottom panel: Decontaminated white dwarf spectrum obtained by subtracting the best-fitting M dwarf template from the observed composite spectrum.}
\vspace{2mm}
\label{fig:fit_M}
\end{figure*}

Figure \ref{fig:fit_M} displays an example of our decontamination method. In general, we obtained excellent fits to the composite spectra, which resulted in very clean white dwarf spectra after subtraction of the M dwarf contribution, as illustrated in the top and bottom panels of Figure \ref{fig:fit_M}. In some cases where the cores of the absorption features are filled in with reprocessed emission from the main-sequence star, we simply excluded the line centers from our fits, as this effect was not taken into account in our binary modeling and thereby could not be corrected for.

\subsection{Problematic Objects} \label{sec:analysis_problem}

Through our visual inspection, we identified 70 DA and 53 (homogeneous) DAO stars affected by the Balmer-line problem. For these objects, the H$\alpha$ and H$\beta$ line cores predicted by our hydrogen-helium models are not as deep as in the observed features, which constitutes an indirect indication of the presence of metals in the atmospheres. Therefore, to improve our atmospheric parameter determination, we re-analyzed these DA and DAO white dwarfs with the CNO-rich models of \citet{gianninas2010}, which generally yielded better fits (see \citealt{gianninas2010} for graphical examples). The most important effect of these models is to increase the surface gravities by $\sim$0.1$-$0.2 dex with respect to those obtained from metal-free models. The Balmer-line problem, found mainly at very high effective temperatures, has an incidence of 34\% among our hydrogen-rich white dwarfs with $\Teff > 60,000$ K. Nonetheless, it is clear that our ability to detect such subtle discrepancies in the line cores depends quite sensitively on the S/N of the spectroscopic data: for instance, the proportion rises to 56\% if we restrict ourselves to S/N $>$ 30. Thus, the Balmer-line problem would likely be more common if we had access to higher-S/N observations. This implies that the atmospheric parameters of some of our very hot DA and DAO stars might suffer from systematic errors. We come back to this point in Section \ref{sec:results} below.

\begin{figure*}
\centering
\includegraphics[width=1.875\columnwidth,clip=true,trim=2.7cm 11.6cm 2.0cm 11.4cm]{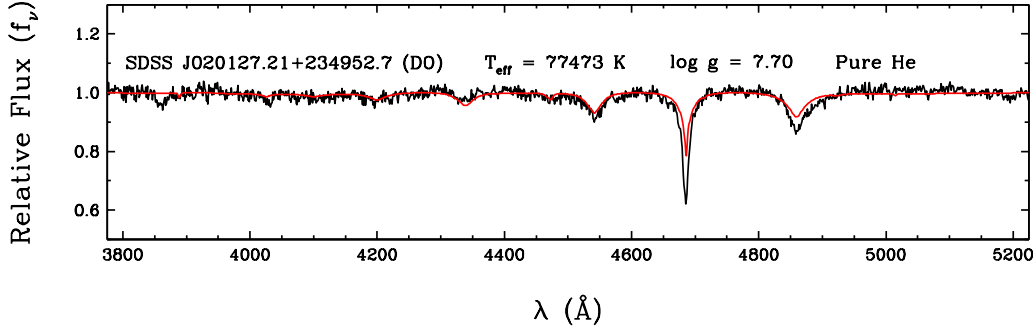}
\caption{Same as Figure \ref{fig:fit_DA}, but for a peculiar DO white dwarf showing abnormally strong He {\sc ii} lines.}
\vspace{2mm}
\label{fig:fit_DOpec}
\end{figure*}

Furthermore, a small number of hot DO and DOZ white dwarfs in our sample are peculiar in that their optical spectra show exceedingly broad and deep He {\sc ii} absorption features that cannot be matched by any of our models. An example of fit for such a strange object is displayed in Figure \ref{fig:fit_DOpec}, where the He {\sc ii} $\lambda$4686 and $\lambda$4859 lines are clearly problematic. In all cases, the issue was already known from previous spectroscopic investigations (see \citealt{reindl2014b} for a compilation). These abnormally strong He {\sc ii} features are often accompanied by so-called ultrahigh-ionization features, believed to be produced by extremely high ionization stages of carbon, nitrogen, oxygen, and neon. For instance, the depression near $\lambda \sim 3850$ \AA\ in the spectrum shown in Figure \ref{fig:fit_DOpec} is attributed to N VII or O VII by \citet{reindl2014b}. The formation of these ions requires temperatures well in excess of 500,000 K, uncharacteristic of a static stellar atmosphere. Hence, the currently favored explanation for this puzzling phenomenon is the presence of a shock-heated circumstellar wind, possibly trapped in a weak magnetic field, around these stars \citep{werner2018,reindl2019}. The reason why such extreme winds occur almost exclusively in helium-dominated white dwarfs still remains a mystery. To assess the atmospheric parameters as well as possible in the circumstances, we rejected the affected He {\sc ii} lines and assigned a higher weight to the He {\sc i} $\lambda$4471, $\lambda$5876, and $\lambda$6678 lines (which persist up to $\Teff \sim 75,000$ K) in our fits. Indeed, while it is uncertain whether the He {\sc ii} features are formed primarily in the hot wind or in the cooler, static atmosphere, the He {\sc i} features must necessarily originate from the latter and can thereby serve as fairly reliable effective temperature indicators. The peculiar DO and DOZ stars all have $\Teff > 60,000$ K and represent 24\% of our helium-rich sample in this temperature range.

\newpage

\section{Results and Discussion} \label{sec:results}

\subsection{Physical Properties} \label{sec:results_param}

For each star in our sample, we converted the atmospheric parameters into stellar parameters through our white dwarf evolutionary sequences\footnote{For the few high-temperature and low-gravity white dwarfs lying outside the parameter space covered by our cooling tracks (see Figure \ref{fig:logg_Teff}), we used a quadratic extrapolation to estimate the stellar properties, which should thus be viewed with caution. Nevertheless, we verified that these extrapolated sequences behave similarly to existing evolutionary calculations, such as those of \citet{althaus2009} and \citet{renedo2010}.}. We used our thick-layer and thin-layer models for hydrogen-rich and hydrogen-deficient objects, respectively. The results are reported in Table \ref{tab:results}, where we give for each white dwarf the SDSS name, spectral type, signal-to-noise ratio (S/N), effective temperature ($\Teff$), surface gravity ($\logg$), atmospheric composition (either pure H, pure He, $\log {\rm He/H}$, or $\log q_{\rm H}$, depending on the adopted chemical structure), mass ($M$), radius ($R$), luminosity ($L$), and cooling age ($\tau_{\rm cool}$)\footnote{Very short cooling ages ($\log \tau_{\rm cool} \ \lta 5$) are sensitive to the zero points set by the initial models of our sequences. In such cases, we state an upper limit ($\log \tau_{\rm cool} < 5$) rather than our derived value.}. The confidence intervals correspond to the internal uncertainties taken directly from the covariance matrix of the fitting procedure. As discussed below, we warn the reader that the derived properties of our hottest objects ($\Teff \ \gta 60,000$ K for DA stars and $\Teff \ \gta 50,000$ K for DO/DB stars) are inaccurate to some degree due to significant systematic effects.

\subsubsection{Examination of the $\logg - \Teff$ Diagram}

\begin{figure*}
\centering
\includegraphics[width=0.975\columnwidth,angle=270,clip=true,trim=5.0cm 2.5cm 4.1cm 1.5cm]{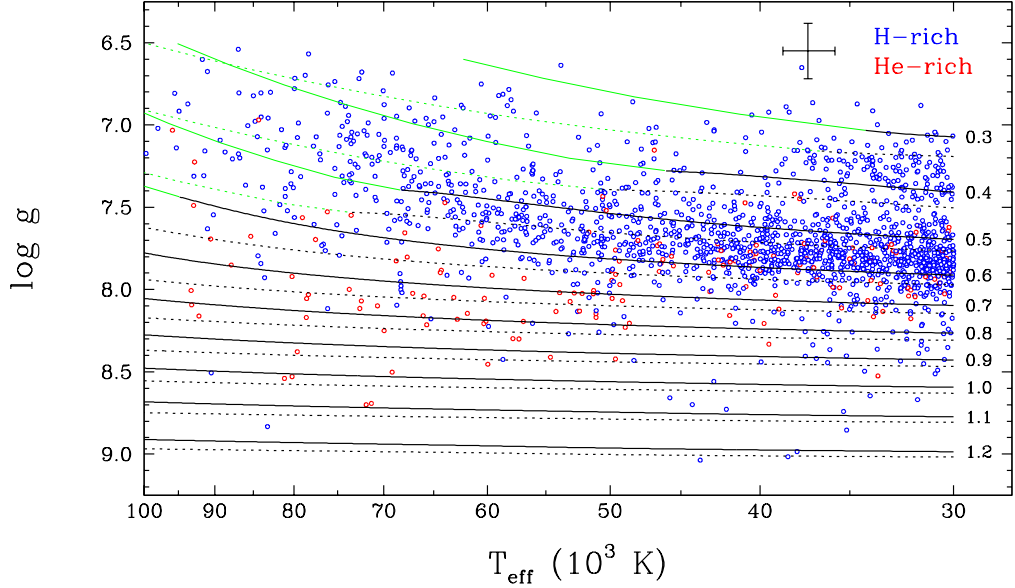}
\caption{Location of our sample of 1806 white dwarfs in the $\logg - \Teff$ diagram, as determined from our spectroscopic analysis. Hydrogen-rich and helium-rich objects are shown in blue and red, respectively. The error bars represent the average uncertainties. Our white dwarf evolutionary sequences for thick and thin hydrogen layers are displayed as solid and dotted black lines, respectively, with masses indicated in units of \msun. The green extensions represent the extrapolated parts of our sequences.} 
\vspace{2mm}
\label{fig:logg_Teff}
\end{figure*}

The atmospheric parameters obtained from our spectroscopic analysis are summarized in the $\logg - \Teff$ diagram presented in Figure \ref{fig:logg_Teff}, where blue and red circles symbolize hydrogen-rich and helium-rich stars, respectively. Also displayed as solid and dotted lines are our cooling tracks for thick and thin hydrogen layers, respectively. On one hand, at low effective temperatures, most DA and DO/DB white dwarfs form a continuous sequence, essentially parallel to the curves of constant mass and tightly centered around $M \sim 0.55$ \msun, as expected from our knowledge of degenerate evolution. We also notice a small population of very low-mass DA stars with $\Teff < 40,000$ K, which are probably remnants of common-envelope binary evolution, since the Galaxy is not old enough for such objects to have been produced through canonical single-star evolution. On the other hand, at high effective temperatures, our results are much more puzzling: DA white dwarfs with $\Teff \ \gta 60,000$ K have lower-than-average surface gravities, and DO white dwarfs with $\Teff \ \gta 50,000$ K have higher-than-average surface gravities. Consequently, the hydrogen-dominated sequence bends upward toward low masses, whereas the helium-dominated sequence abruptly drops toward high masses. These features are definitely at odds with the well-established facts that white dwarfs evolve at constant mass and that the DA and DB spectral classes are characterized by similar mean masses \citep{ourique2019,tremblay2019b,genest-beaulieu2019a}.

\begin{figure*}
\centering
\includegraphics[width=0.975\columnwidth,clip=true,trim=2.4cm 4.8cm 2.2cm 5.2cm]{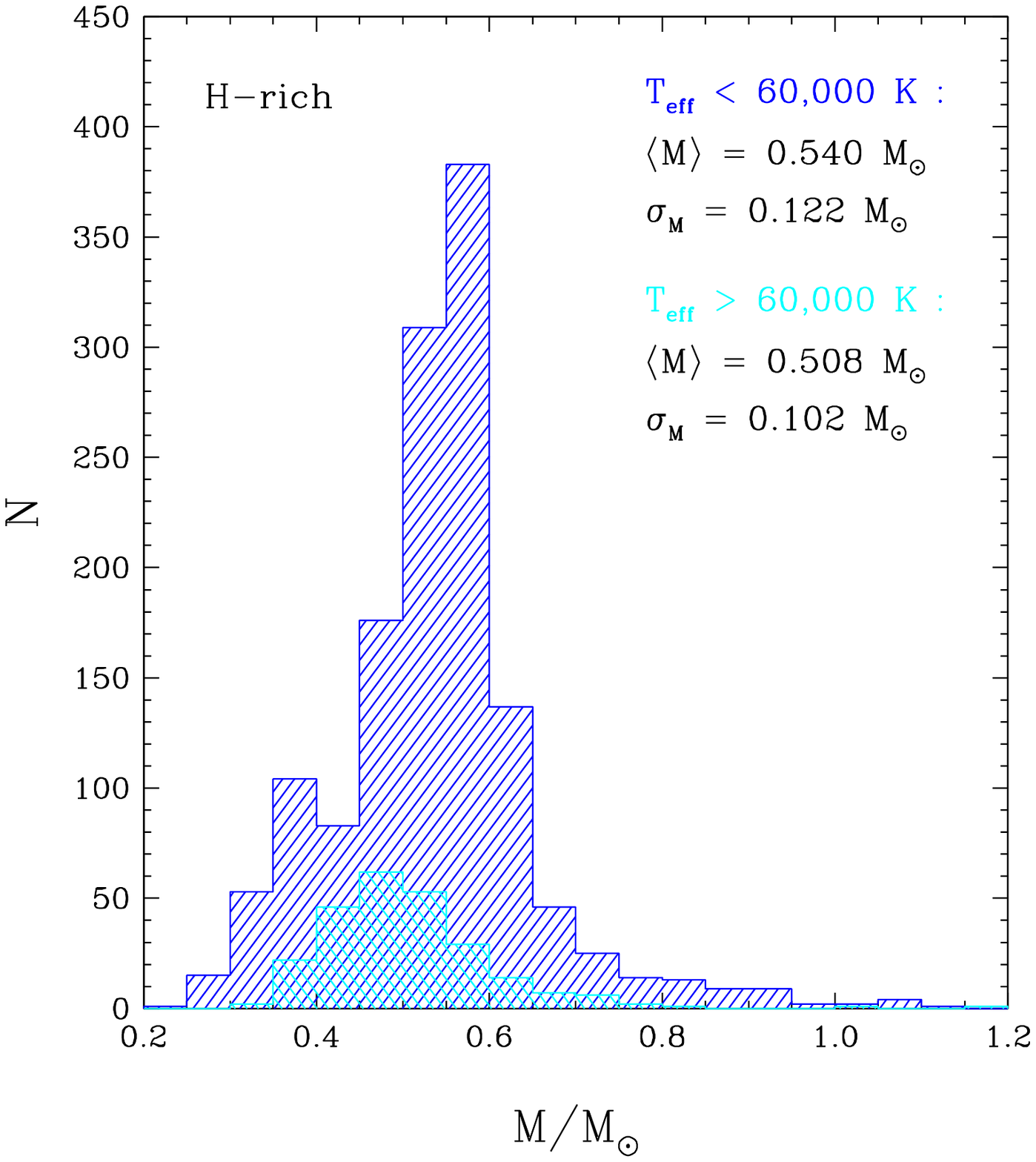}
\includegraphics[width=0.975\columnwidth,clip=true,trim=2.8cm 4.8cm 1.8cm 5.2cm]{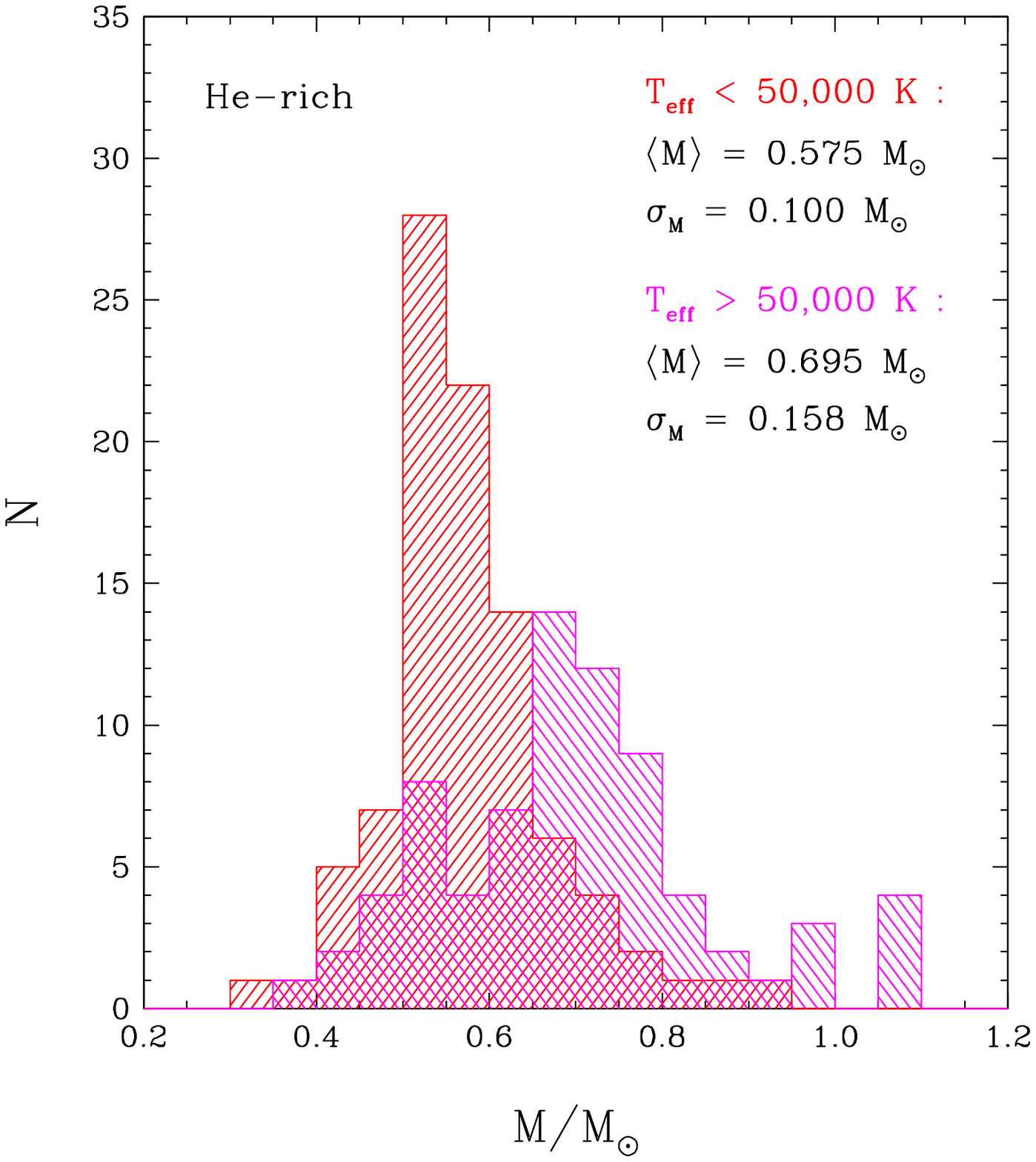}
\caption{Cumulative mass distributions of hydrogen-rich (left panel) and helium-rich (right panel) objects in our sample of 1806 white dwarfs, as determined from our spectroscopic analysis. Each group is divided into a low-temperature subgroup (blue and red histograms in the left and right panels, respectively) and a high-temperature subgroup (cyan and magenta histograms in the left and right panels, respectively) based on the trends seen in Figure \ref{fig:logg_Teff}. The average and standard deviation of the various distributions are given in the panels.}
\vspace{2mm}
\label{fig:N_M}
\end{figure*}

The problem is illustrated more quantitatively in Figure \ref{fig:N_M}, where we show the cumulative mass distributions for hydrogen-atmosphere and helium-atmosphere objects. Each group is divided into a low-temperature sample and a high-temperature sample (based on Figure \ref{fig:logg_Teff}, the boundary is set at $\Teff = 60,000$ K for DA stars and at $\Teff = 50,000$ K for DO/DB stars). As anticipated from Figure \ref{fig:logg_Teff}, the mass distribution of the hottest DA white dwarfs is slightly shifted toward lower masses with respect to that of their cooler counterparts. Accordingly, the mean mass is slightly lower at $\Teff > 60,000$ K ($\langle M \rangle = 0.508$ \msun) than at $\Teff < 60,000$ K ($\langle M \rangle = 0.540$ \msun). For helium-rich objects, the discrepancy between the high-temperature and low-temperature regimes is even more severe: the mass distribution of hot DO stars is flatter and displaced to considerably higher masses. This translates into significantly larger values of the average and the standard deviation at $\Teff > 50,000$ K ($\langle M \rangle = 0.695$ \msun) than at $\Teff < 50,000$ K ($\langle M \rangle = 0.575$ \msun). On a more encouraging note, the low-temperature DA and DO/DB mass distributions are comparable (with the exception of a small low-mass peak in the DA histogram associated with post$-$common-envelope white dwarfs), as expected. We are thus confident that our gravity and mass determinations are relatively reliable in this temperature range, where the bulk of our sample lies.

It should be mentioned that the mass issues raised here are not unique to our study. In their recent spectroscopic analyses of SDSS DA white dwarfs, \citealt{kepler2019} (see their Figure 2) and \citealt{genest-beaulieu2019a} (see their Figure 7) also found a trend of decreasing mass with increasing effective temperature at the hot end of their samples. Furthermore, from their compilation of spectroscopic parameters of SDSS DO stars, \citealt{reindl2014b} (see their Figure 5) also obtained a mass distribution with a peak at $M \sim 0.675$ \msun. They argue that this high-mass peak is real, but the striking contrast between the DO and DB mass distributions suggests otherwise. Therefore, it seems clear that the spectroscopic masses of very hot SDSS white dwarfs are afflicted by generalized problems that are not restricted to our own model atmospheres, cooling tracks, or fitting technique.

\subsubsection{Insight from Photometric and Astrometric Data}

Our assertion that the high-temperature mass offsets are artificial can be proven by deriving alternative mass estimates through an independent method. This is where the $ugriz$ photometry and {\it Gaia} parallaxes can be useful. Indeed, the photometric energy distribution of a white dwarf principally depends on its effective temperature $\Teff$ and solid angle $\pi(R/D)^2$, and thereby on its radius $R$ (and mass $M$ via the mass-radius relation) if the distance $D$ is known from a parallax measurement \citep{bergeron1997,bergeron2001}. For each of the 824 objects in our astrometric subsample ($\sigma_\pi/\pi \le 25\%$), we transformed the dereddened and SDSS-to-AB corrected $ugriz$ magnitudes into observed average fluxes, which we then fitted with theoretical average fluxes predicted from our model atmospheres using the Levenberg-Marquardt algorithm. In the fitting procedure, only the radius was treated as a free parameter, while we assumed the effective temperature and atmospheric composition obtained from our spectroscopic analysis. This approach was motivated by the fact that the optical energy distribution of hot stars is very weakly sensitive to effective temperature, as shown in Section \ref{sec:sample}. For white dwarfs with main-sequence companions, the latter often contribute considerably to the measured $riz$ magnitudes (and sometimes to the $g$ magnitude as well), which were consequently excluded from the fits. Finally, we employed our evolutionary sequences to convert the inferred radius into surface gravity and mass, henceforth denoted as $\logg_\pi$ and $M_\pi$.

\begin{figure*}
\centering
\includegraphics[width=0.975\columnwidth,angle=270,clip=true,trim=5.0cm 2.5cm 4.1cm 1.5cm]{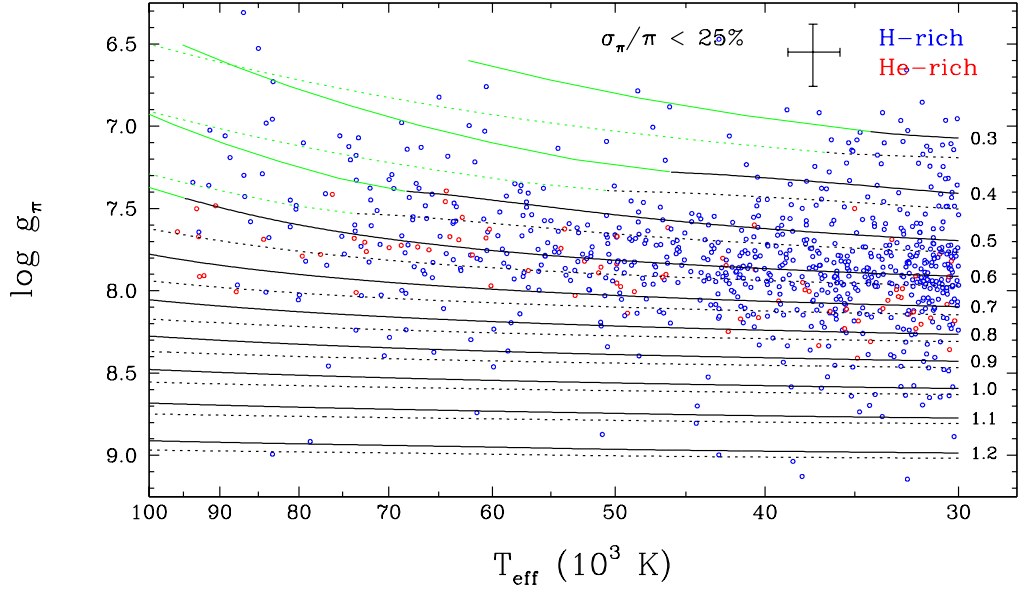}
\caption{Same as Figure \ref{fig:logg_Teff}, but for our subsample of 824 white dwarfs with $\sigma_\pi/\pi \le 25\%$, and as determined from our photometric analysis.} 
\vspace{2mm}
\label{fig:loggpi_Teff}
\end{figure*}

\begin{figure*}
\centering
\includegraphics[width=0.975\columnwidth,clip=true,trim=2.4cm 4.8cm 2.2cm 5.2cm]{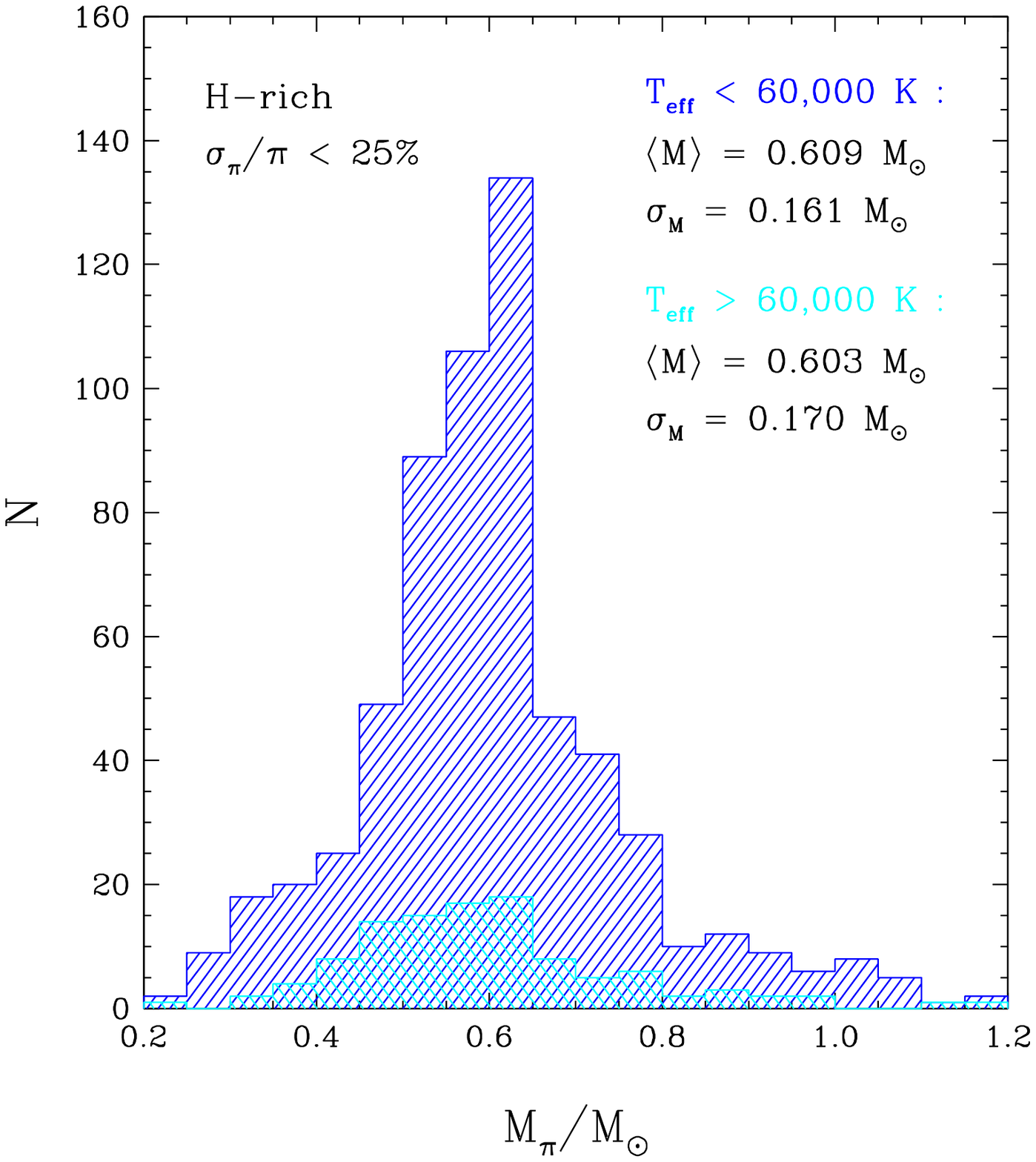}
\includegraphics[width=0.975\columnwidth,clip=true,trim=2.8cm 4.8cm 1.8cm 5.2cm]{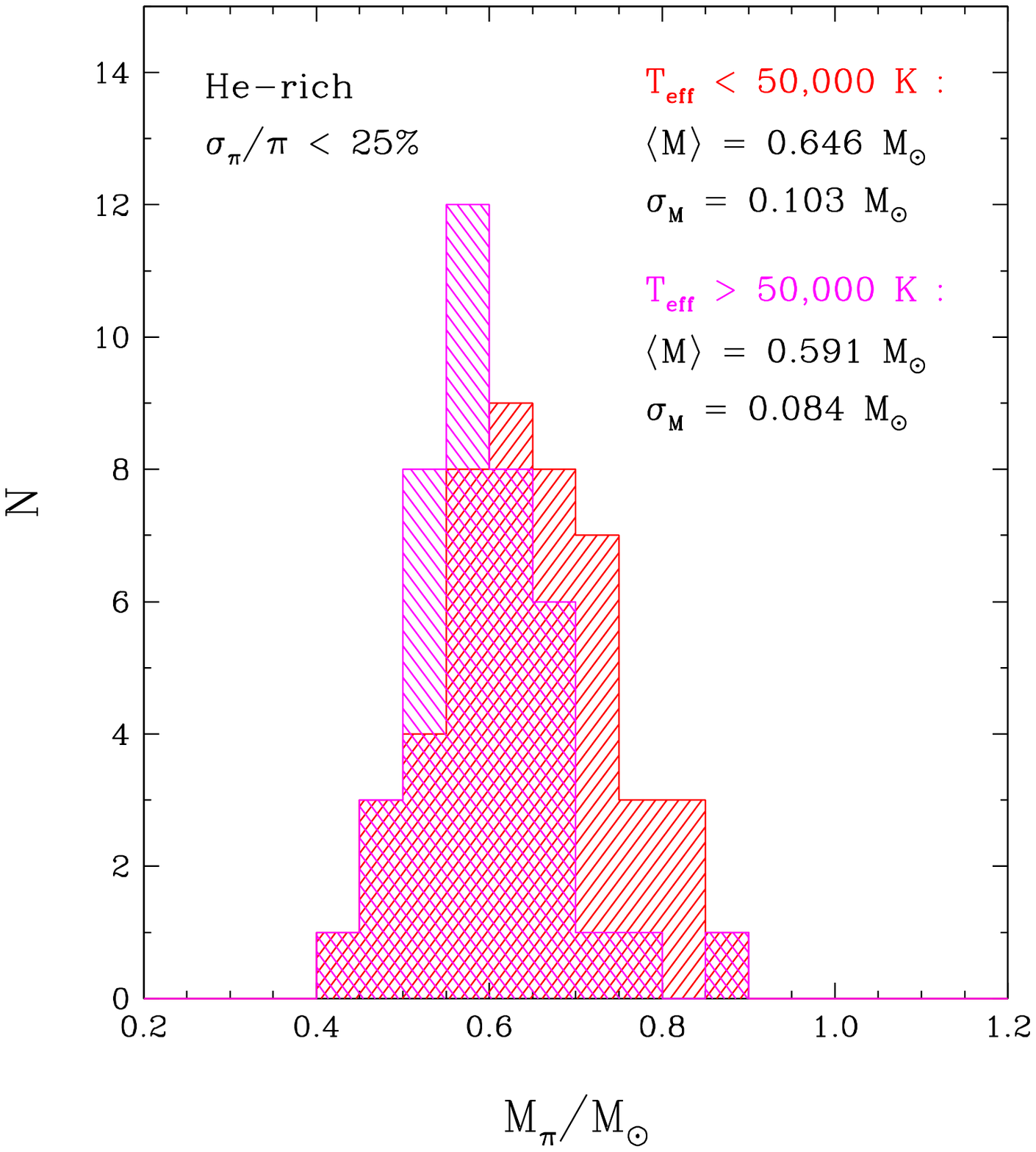}
\caption{Same as Figure \ref{fig:N_M}, but for our subsample of 824 white dwarfs with $\sigma_\pi/\pi \le 25\%$, and as determined from our photometric analysis.}
\vspace{2mm}
\label{fig:N_Mpi}
\end{figure*}

The results of this photometric analysis are presented in Figures \ref{fig:loggpi_Teff} and \ref{fig:N_Mpi}, which are exactly analogous to Figures \ref{fig:logg_Teff} and \ref{fig:N_M}, but with $\logg_\pi$ and $M_\pi$ in place of $\logg$ and $M$. This time, the hot DA and DO white dwarfs form a smooth, constant-mass extension of the low-temperature cooling sequence in the $\logg - \Teff$ diagram. Accordingly, the high-temperature and low-temperature mass distributions are now in markedly better agreement, for both hydrogen-atmosphere and helium-atmosphere objects. We still notice a small average offset of $\sim$0.05 \msun\ between the hot and cool helium-rich samples, but the corresponding histograms now largely overlap and have similar shapes, two notable improvements over the spectroscopic case. On another note, comparing the photometric and spectroscopic mass distributions in the low-temperature regime (where the reliability of our atmospheric parameters is likely highest), it can be seen that the two mass scales are shifted by $\sim$0.07 \msun\ with respect to each other, for both DA and DO/DB stars. This disparity was already exposed in previous investigations of SDSS white dwarfs and was attributed to problems in the calibration of the SDSS spectra causing the surface gravities measured spectroscopically to be slightly too low \citep{tremblay2011,genest-beaulieu2014,genest-beaulieu2019a}.

\subsubsection{Possible Error Sources}

The above test clearly demonstrates that the systematic shifts observed at the hot end of our spectroscopic $\logg - \Teff$ diagram are not real. Then, what is wrong with the spectroscopic gravities and masses of very hot white dwarfs? The answer to this question is certainly not identical for DA and DO stars, since we found opposite trends for these two groups. Nevertheless, in both instances, the issue must be related to the spectral line profiles, as we showed that the overall energy distribution is not affected.

In the case of DA white dwarfs, valuable insight can be gained from a comparison with the work of \citet{gianninas2010}, who contrarily obtained normal surface gravities (see their Figure 14). There are two primary differences between their study and ours. First, they conducted their own spectroscopic survey and hence did not rely on SDSS observations. Second, thanks to their high-S/N spectra, they detected the Balmer-line problem in a significant fraction of their objects with $\Teff > 60,000$ K, which they consequently analyzed with CNO-rich model atmospheres. We believe that both our use of SDSS data and our use of metal-free models contributed to the mass shift encountered here. First, as mentioned above, it is now well-documented that the SDSS spectra suffer from a small calibration issue, which leads to slightly underestimated $\logg$ values. This effect is probably aggravated at higher temperatures by the lower sensitivity of the H {\sc i} line profiles to the atmospheric parameters. Second, as discussed in Section \ref{sec:analysis}, only 34\% of our DA and DAO stars with $\Teff > 60,000$ K exhibit the Balmer-line problem, but this proportion rises to 56\% if we only consider objects with S/N $>$ 30. Therefore, the presence of small amounts of atmospheric metals might be rather frequent among our hottest hydrogen-dominated white dwarfs, in line with the conclusions of FUV studies \citep{barstow2003a,barstow2014,good2005}. Yet the corresponding Balmer-line problem sometimes remains imperceptible, buried in the noise of the SDSS data, in which situation our analysis relying on hydrogen-helium models likely underevaluated the surface gravities. To further investigate this idea, we performed an experiment in which we fitted all of our hot DA and DAO spectra with the CNO-rich models of \citet{gianninas2010}. The outcome was that the high-temperature mean mass increased to $\langle M \rangle = 0.520$ \msun, closer to (but still a bit lower than) the low-temperature mean mass of $\langle M \rangle = 0.540$ \msun, in accordance with our expectations.

In the case of DO white dwarfs, the significantly too high $\logg$ values point to a severe problem in the modeling of the He {\sc ii} line broadening. A simple explanation would be that the Stark broadening profiles of \citet{schoening1989}, used in all modern analyses of DO stars, are inadequate for some unknown reason. There is, however, another conceivable possibility. As mentioned in Section \ref{sec:analysis}, a non-negligible fraction of the hottest helium-atmosphere white dwarfs show extremely broad and deep He {\sc ii} features stemming from an ultrahot circumstellar wind \citep{werner2018,reindl2019}. Given our results, it is tempting to generalize this concept and to suggest that most DO stars with $\Teff \ \gta 50,000$ K are afflicted by this phenomenon, albeit to a lower degree. The wind would then act as an additional line broadening mechanism, obviously not included in our static model-atmosphere calculations and thus inducing a spuriously high surface gravity. Still, this interpretation seems a little far-fetched, and we believe that the current modeling of Stark broadening of He {\sc ii} features should be looked into first.

Given the above considerations about the surface gravities, one may legitimately ask, how trustworthy are the effective temperatures? This question is of fundamental importance for our forthcoming discussion of the spectral evolution of hot white dwarfs, which is largely based on our spectroscopically derived $\Teff$ values. Several works have cast doubts on the accuracy of the temperature scale of very hot white dwarfs as measured from optical spectroscopy (\citealt{barstow1998b,barstow2003b,good2004,werner2017,werner2019}; see also \citealt{latour2015} in the context of hot subdwarfs). Accordingly, we recognize that our effective temperatures probably suffer from sizable systematic errors in the range $\Teff \ \gta 75,000$ K, a caveat that we bear in mind when assessing the spectral evolution below. Nevertheless, we have good reasons to trust our effective temperatures in the range $\Teff \ \lta 75,000$ K. In the case of DA stars, a comparison of Figures \ref{fig:logg_Teff} and \ref{fig:loggpi_Teff} indicates that the $\logg$ offset remains rather small and thus that our spectroscopic solutions are only marginally affected in this temperature regime. Moreover, while switching from CNO-free to CNO-rich models causes conspicuous changes in $\logg$, the corresponding changes in $\Teff$ are typically minor (see Figure 11 of \citealt{gianninas2010}). In the case of DO/DB stars, even if the $\logg$ shift is substantially worse, the effective temperatures are still robustly constrained by the simultaneous presence of He {\sc i} and He {\sc ii} features in the optical spectra for $\Teff \ \lta 75,000$ K. This is because the ionization balance of helium, which determines the relative strength of the two sets of lines, is strongly temperature-dependent and weakly gravity-dependent. As an illustrative experiment, if we fit a $\Teff = 65,000$ K, $\logg = 7.8$, pure-helium synthetic spectrum with our full grid of pure-helium synthetic spectra while forcing the surface gravity to the much higher value $\logg = 8.2$, the resulting effective temperature is $\Teff = 65,730$ K, only 1.1\% higher than the true value. In short, despite the shortcomings of our analysis, we are confident that our spectroscopic temperature scale is reliable up to $\Teff \sim 75,000$ K.

\subsubsection{Parameters of Hybrid White Dwarfs}

\begin{figure*}
\centering
\includegraphics[width=0.975\columnwidth,angle=270,clip=true,trim=5.0cm 2.5cm 4.1cm 1.5cm]{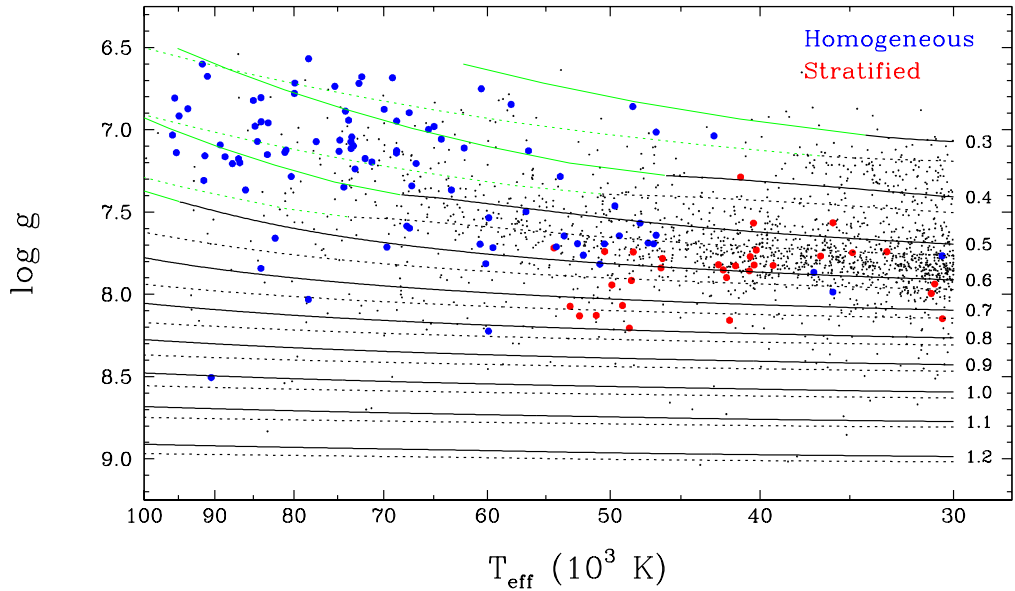}
\caption{Same as Figure \ref{fig:logg_Teff}, but with emphasis on the 125 single hybrid white dwarfs in our sample (we exclude the two double degenerate systems mentioned in Section \ref{sec:analysis}). Objects with chemically homogeneous and stratified atmospheres are shown as large blue and red circles, respectively. For comparison purposes, the rest of the stars in our sample are represented as small black dots.} 
\vspace{2mm}
\label{fig:logg_Teff_hyb}
\end{figure*}

We end this subsection by addressing the properties of the hybrid white dwarfs in our sample. The location of these stars in the $\logg - \Teff$ diagram is highlighted in Figure \ref{fig:logg_Teff_hyb}, where chemically homogeneous and stratified atmospheres are shown in blue and red, respectively. On one hand, the bulk of our homogeneous objects have hydrogen-rich atmospheres containing small traces of helium and populate the high-temperature, low-gravity part of the $\logg - \Teff$ diagram. Those belong to the group of classical DAO white dwarfs, which have been known for a long time and extensively studied over the years \citep{bergeron1994,napiwotzki1999,good2005,gianninas2010,tremblay2011,kepler2019}. Similarly to our hot DA stars, the surface gravities and masses of our hot homogeneous DAO stars are obviously underestimated. Nonetheless, we can still deduce from Figure \ref{fig:logg_Teff_hyb} that these objects truly have mildly low masses in a relative sense (that is, compared to DA stars of the same temperature), a trend also noticed in most papers just cited. On the other hand, the stratified white dwarfs, which were discovered in significant numbers only recently by \citet{manseau2016}, are located in a completely different part of the $\logg - \Teff$ diagram: they all have $\Teff < 55,000$ K and more typical masses. Furthermore, as stated in Section \ref{sec:analysis}, they are characterized by hydrogen layer masses in the full range $-18.25 < \log q_{\rm H} < -15.25$ and hence exhibit all varieties of spectra from helium-line$-$dominated (spectral type DOA/DBA) to hydrogen-line$-$dominated (spectral type DAO/DAB). This evidence strongly supports the idea that homogeneous and stratified white dwarfs form two fundamentally distinct populations, as pointed out by \citet{manseau2016}.

\subsection{Spectral Evolution} \label{sec:results_evol}

We are now ready to address the central purpose of the paper, which is to improve our understanding of the spectral evolution of hot white dwarfs. To do so, we rely essentially on the effective temperatures and atmospheric compositions inferred from our spectroscopic analysis. Because we want our picture of the spectral evolution to be as faithful as possible, from now on we only consider the objects with S/N $\ge$ 10 (1467 out of 1806 stars), a restriction that offers a good compromise between the size and quality of the ensuing sample.

\subsubsection{Selection Bias Corrections}

Ideally, one could study the spectral evolution by simply counting the numbers of hydrogen-atmosphere and helium-atmosphere objects in several effective temperature bins and examining how these numbers change along the white dwarf cooling sequence. However, in the present work, there are two potentially important selection effects that might render this approach inadequate and therefore must be discussed before going further. 

First, a possible variation of the SDSS spectroscopic completeness with temperature and/or composition could seriously distort the results. Fortunately for us, this is not an issue for our sample. By virtue of their very blue colors, most of our objects were likely observed as part of the so-called ``hot-standard'' target class of the SDSS, independently of their specific atmospheric properties \citep{eisenstein2006b}. However, it is important to note that such a bias was actually induced by our own sample selection process. Indeed, it is well-known that the hydrogen-deficient white dwarf population comprises not only the helium-rich DO stars, but also the helium-, carbon-, and oxygen-rich PG 1159 stars at the very hot end of the cooling sequence ($\Teff \ \gta 80,000$ K; \citealt{werner2006}). In our initial color-selected sample (see Section \ref{sec:sample}), we identified 17 PG 1159 stars but did not retain them in our final sample given that our hydrogen-helium model atmospheres are not appropriate for their analysis. In the following, we take these 17 objects into account by using atmospheric parameters from the literature \citep{dreizler1998,nagel2006,hugelmeyer2006,werner2014,kepler2016}.

\begin{figure}
\centering
\includegraphics[width=0.975\columnwidth,clip=true,trim=2.3cm 7.6cm 1.8cm 6.0cm]{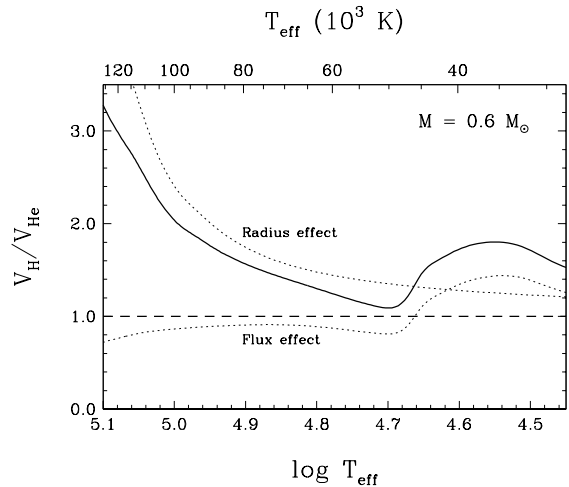}
\caption{Ratio of the volumes of space sampled by hydrogen-atmosphere and helium-atmosphere white dwarfs (with $M = 0.6$ \msun) in a $g$-band magnitude-limited survey as a function of effective temperature. The dotted lines show the individual contributions of the flux and radius differences to the total ratio.}
\vspace{2mm}
\label{fig:volcorr}
\end{figure}

Second, since the SDSS is a magnitude-limited survey and since intrinsically brighter stars are more preferentially detected, hotter white dwarfs are inevitably over-represented in our sample. This selection effect can be partly eliminated by working with fractions or ratios of objects rather than with absolute numbers, because at a given temperature white dwarfs of different surface compositions have similar luminosities and are thus affected to a comparable extent by this bias. Consequently, we choose here to discuss the spectral evolution in terms of fractions of stars of some types (for instance, the fraction of helium-atmosphere white dwarfs). Still, these fractions must be corrected for the fact that hydrogen-rich and helium-rich white dwarfs with identical atmospheric parameters do not have strictly equal luminosities. This is because they do not have the same emergent fluxes (due to their different atmospheric opacities) and radii (due to their different hydrogen layer masses). To account for this phenomenon, we rely on the so-called maximum-volume method \citep{schmidt1975}. More specifically, we estimate the volumes $V_{\rm H}$ and $V_{\rm He}$ corresponding to the maximum distances at which DA and DO/DB stars of a given temperature would still be observed in a $g$-band magnitude-limited survey such as the SDSS. We use pure-hydrogen atmospheres and thick-layer cooling tracks for the calculation of $V_{\rm H}$, whereas we use pure-helium atmospheres and thin-layer cooling tracks for the calculation of $V_{\rm He}$. We also simply assume $M = 0.6$ \msun\ throughout. The ratio $V_{\rm H}/V_{\rm He}$ of the volumes probed by each type of objects represents the quantitative factor by which the detection rates of hydrogen-dominated and helium-dominated white dwarfs are expected to differ. Figure \ref{fig:volcorr} shows this ratio, as well as the individual contributions of the flux and radius effects mentioned above, as a function of effective temperature. The combination of both effects leads to $V_{\rm H}/V_{\rm He} > 1$ over the entire temperature range considered here, meaning that a DA star is always brighter and hence seen up to a farther distance than an analogous DO/DB star. To compensate for this bias, we assign a modified weight of $V_{\rm H}/V_{\rm He}$ to each helium-rich white dwarf, as opposed to a weight of 1 to each hydrogen-rich white dwarf, in our statistical analysis. See \citet{eisenstein2006a}, \citet{tremblay2008}, and \citet{blouin2019} for similar applications of this method.

\subsubsection{Constraints from Helium-rich White Dwarfs}

\begin{figure}
\centering
\includegraphics[width=0.975\columnwidth,clip=true,trim=2.3cm 4.9cm 1.8cm 3.3cm]{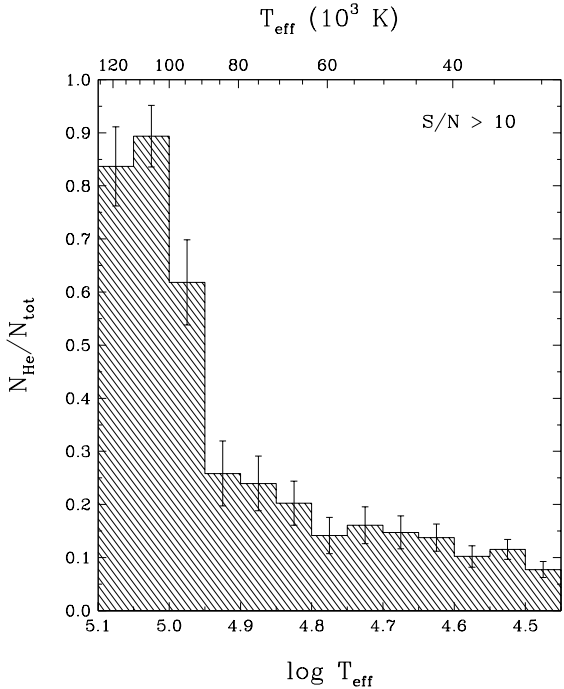}
\caption{Fraction of helium-rich white dwarfs (corrected for the selection effects discussed in the text) as a function of effective temperature for our subsample of 1467 objects with S/N $\ge$ 10. The error bars represent the Poisson statistics of each bin.}
\vspace{2mm}
\label{fig:f_Teff_DO}
\end{figure}

Figure \ref{fig:f_Teff_DO} presents the fraction of helium-rich objects (corrected for the selection effects described above) as a function of decreasing effective temperature. Focusing first on the high-temperature part of the histogram, we notice that as much as $\sim$85$-$90\% of white dwarfs have hydrogen-deficient atmospheres at the very beginning of the cooling sequence. Then, this proportion drops sharply to $\sim$25\% below $\Teff \sim 90,000$ K. This glaring lack of extremely hot DA stars is a long-standing problem of spectral evolution theory \citep{fleming1986,krzesinski2009,werner2019}. Taken at face value, this result seems to imply that the vast majority of all white dwarfs are initially born with helium-rich atmospheres, but then develop hydrogen-rich atmospheres quite early in their evolution through some mechanism, as originally suggested by \citet{fontaine1987}. Nonetheless, this interpretation is rather doubtful for two reasons. First, as mentioned earlier, the temperature scale of these very hot objects suffer from significant uncertainties, which certainly impact the statistics reported in Figure \ref{fig:f_Teff_DO}, although to an unclear extent. Second, as proposed many years ago by \citet{fleming1986} and more recently by \citet{werner2019}, the deficiency of DA white dwarfs with $\Teff > 90,000$ K might be an artifact of the different evolutionary rates of hydrogen-dominated and helium-dominated objects. Indeed, at the entrance of the cooling sequence, nuclear burning switches off more rapidly in hydrogen envelopes than in helium envelopes, causing the former to sustain a high luminosity for a shorter time than the latter \citep{iben1984}. Therefore, it is entirely plausible that the apparent paucity of extremely hot hydrogen-rich white dwarfs is not due to some kind of atmospheric transformation, but rather to the fact that DA stars pass through this high-temperature phase faster than their DO (and PG 1159) counterparts\footnote{According to this argument, our implicit assumption that the effective temperature is a good proxy for the cooling age in our discussion of the spectral evolution, although valid over most of the cooling sequence, fails for $\Teff > 90,000$ K. However, using the cooling age as the independent variable is not a viable alternative, for several reasons: (1) our evolutionary tracks do not include residual nuclear burning and the associated difference in cooling rate between models with thick and thin hydrogen layers; (2) for very young white dwarfs ($\log \tau_{\rm cool} \ \lta 5$), our cooling ages are sensitive to the zero points defined by the initial models of our tracks; (3) since $\tau_{\rm cool}$ depends on both $\Teff$ and $\logg$, the systematic errors on our surface gravities cast doubts on the reliability of our cooling ages at very high effective temperatures.}. We note in passing that such a difference in cooling timescale between hydrogen-envelope and helium-envelope pre-white dwarfs is actually predicted by the evolutionary calculations of \citet{althaus2009} and \citet{renedo2010}, as shown in Figure 11 of \citet{werner2019}. In the end, we simply refrain from drawing any firm conclusion on the spectral evolution based on our results at very high effective temperatures.

Fortunately, the statistics for cooler stars are not subject to these difficulties. As argued earlier, for $\Teff \ \lta 75,000$ K, our temperature scale rests on more solid grounds. Moreover, at these temperatures, the cooling ages of hydrogen-rich and helium-rich white dwarfs become similar, as the longer evolutionary timescale makes the initial difference completely negligible. Thus, the fraction of helium-dominated objects at $\Teff \sim 75,000$ K, which is $\sim$24\% according to Figure \ref{fig:f_Teff_DO}, accurately reflects the actual population of very hot white dwarfs. In other words, we can confidently assert that $\sim$24\% of all white dwarfs are born with hydrogen-deficient atmospheres.

Going down in effective temperature, we see in Figure \ref{fig:f_Teff_DO} that the proportion of helium-rich white dwarfs gradually declines from $\sim$24\% at $\Teff \sim 75,000$ K to $\sim$8\% at $\Teff \sim 30,000$ K. In that respect, our results smoothly merge with those of \citealt{ourique2019} (see their Figure 10) and \citealt{genest-beaulieu2019b} (see their Figure 23) at lower temperatures. Unlike the sharp drop at $\Teff \sim 90,000$ K, there are no reasons for this continuous decrease not to be real, which suggests that spectral evolution does occur here. More specifically, among the $\sim$24\% of DO stars at $\Teff \sim 75,000$ K, $\sim$1/3 remain helium-rich while $\sim$2/3 become hydrogen-rich by the time they reach $\Teff \sim 30,000$ K. From the results of \citet{ourique2019} and \citet{genest-beaulieu2019b}, we can even argue that those persisting DB stars at $\Teff \sim 30,000$ K will always remain so at lower temperatures, because the fraction of helium-atmosphere white dwarfs does not decrease further for $\Teff < 30,000$ K (it stays roughly constant down to $\Teff \sim 20,000$ K and then increases). Our findings indicate that the concept of a DB gap, historically seen as a hole with sharp boundaries, must be definitely abandoned: this so-called gap not only contains a few DB stars, as first discovered by \citet{eisenstein2006a}, but also has a fuzzy blue edge, since the deficiency of helium-rich white dwarfs grows progressively over a very broad range of effective temperatures ($75,000 \ {\rm K} > \Teff > 30,000$ K).

The most likely physical explanation for the transformation of DO stars into DA stars is provided by the so-called float-up model of \citet{fontaine1987}. This scenario begins with a PG 1159 star, whose peculiar surface chemistry is the result of a late helium-shell flash, during which the superficial hydrogen is deeply mixed in the helium envelope and almost entirely burned, while significant quantities of carbon and oxygen are dredged up to the surface \citep{herwig1999,althaus2005,werner2006}. The basic assumption of \citet{fontaine1987} is that a small amount of hydrogen survives this event, but initially remains undetectable, since it is diluted in the whole envelope. As gravitational settling starts to operate, carbon and oxygen rapidly sink out of the atmosphere, and the PG 1159 star becomes a DO white dwarf. Then, the residual hydrogen slowly diffuses upward and accumulates at the surface, building up a thicker and thicker superficial hydrogen layer, and eventually turning the DO star into a DA star.

Although the float-up model offers a relevant description of the DO-to-DA transition, we stress that some aspects of the original proposition of \citet{fontaine1987} no longer hold today. Based on the absence of DA stars with $\Teff > 80,000$ K and of DB stars with $45,000 \ {\rm K} > \Teff > 30,000$ K in the PG survey, they argued that the entire white dwarf population descends from PG 1159 stars and undergoes the float-up of residual hydrogen. We now know that this is not the case. First, as discussed above, the deficiency of extremely hot hydrogen-rich white dwarfs is probably only apparent, and the true fraction of helium-rich white dwarfs at the beginning of the cooling sequence is likely closer to $\sim$24\%. Second, the PG 1159 stars are not the sole precursors of DO stars. In fact, the helium-dominated white dwarf population is also fed by a second evolutionary channel associated with a different progenitor: the so-called O(He) stars, which display almost pure-helium atmospheres and are believed to be formed through merger or common-envelope processes \citep{rauch1998,reindl2014b,reindl2014a}. Third, not all DO stars evolve into DA stars: $\sim$8\% of all white dwarfs never become hydrogen-rich at any point during their life.

In the framework of the float-up model, the results presented in Figure \ref{fig:f_Teff_DO} can be interpreted as follows. Among the DO stars at $\Teff \sim 75,000$ K, $\sim$2/3 (or $\sim$16\% of the whole white dwarf population) contain residual hydrogen that diffuses to the surface and turns them into DA stars before they reach $\Teff \sim 30,000$ K, whereas the remaining $\sim$1/3 (or $\sim$8\% of the whole white dwarf population) do not possess a sufficient amount of hydrogen for this transformation to happen. The fact that the fraction of helium-dominated objects declines very gradually with decreasing effective temperature then suggests that a broad range of total hydrogen content exists within the DO/DB population. Qualitatively, the larger the quantity of residual hydrogen, the faster the float-up process, and the earlier the helium-to-hydrogen atmospheric transition. Quantitatively, estimating total hydrogen masses from our results would require detailed time-dependent simulations of diffusion in evolving white dwarfs. Indeed, when the spectral type of a star changes from DO to DA, it merely means that enough hydrogen has accumulated at the surface so as to enclose the full line-forming region, which corresponds to a lower limit of $q_{\rm H} \ \gta 10^{-15}$. It is almost certain that much more hydrogen is still hidden deeper in the envelope, where the diffusion timescales are extremely long and the composition profile thereby remains far from diffusive equilibrium \citep{dehner1995,althaus2004,rolland2020}. This hydrogen reservoir, which can only be probed through theoretical calculations of gravitational settling, might play a fundamental role in the spectral evolution of cooler white dwarfs \citep{rolland2020}.

As for the objects that always retain helium-rich atmospheres, there is simply no way to tell whether they represent the tail of the $q_{\rm H}$ distribution and thus contain hydrogen in an amount too small to ever be detected, as alleged by \citet{koester2015}, or arise from a distinct evolutionary path that made them completely devoid of hydrogen, as claimed by \citet{bergeron2011}. Nevertheless, it is interesting to note that our two groups of DO white dwarfs (those that eventually become hydrogen-rich and those that do not) are somewhat reminiscent of the two known formation channels of DO white dwarfs, involving respectively the PG 1159 and O(He) stars. It is not unreasonable to speculate that the two types of progenitors lead to two different spectral evolution scenarios, with and without residual hydrogen.

It is worth mentioning that several works have attempted to infer the total hydrogen content of cooler degenerates from measured atmospheric compositions \citep{macdonald1991,tremblay2008,koester2015,rolland2018,rolland2020,genest-beaulieu2019b,cunningham2020,koester2020}. Briefly, the increase in the fraction of helium-atmosphere objects for $20,000 \ {\rm K} > \Teff > 6000$ K implies hydrogen masses in the range $10^{-16} < q_{\rm H} < 10^{-8}$ for $\sim$15$-$30\% of all white dwarfs. If a DA star is characterized by $q_{\rm H} \ \lta 10^{-14}$, it will transform into a DB star as a consequence of the so-called convective dilution process, by which the thin hydrogen layer is eroded from below by the much thicker convective helium envelope. Otherwise, a DA white dwarf still has the opportunity to develop a helium-rich atmosphere when the superficial hydrogen layer in turn becomes convective and mixes with the underlying helium mantle, a phenomenon referred to as convective mixing. The thicker the hydrogen layer, the lower the transition temperature, hence a range of $q_{\rm H}$ values translates into a range of $\Teff$ values over which white dwarfs experience convective dilution or mixing. However, the quantitative results cited above must be taken with caution, since they were obtained under the questionable assumption that all the hydrogen has had enough time to float to the surface. Once again, we emphasize that this is likely not the case, given the very long diffusion timescales at the base of white dwarf envelopes. In fact, the parameter $q_{\rm H}$ reported in the literature should be viewed as the amount of hydrogen residing in the outer envelope at a given time, and not as the total amount of hydrogen present in the star. Because of this widespread confusion, we refrain from further discussing the matter until the evolution of the inner chemical structure is better understood.

\subsubsection{Constraints from Hybrid White Dwarfs}

\begin{figure}
\centering
\includegraphics[width=0.975\columnwidth,clip=true,trim=2.3cm 4.9cm 1.8cm 3.3cm]{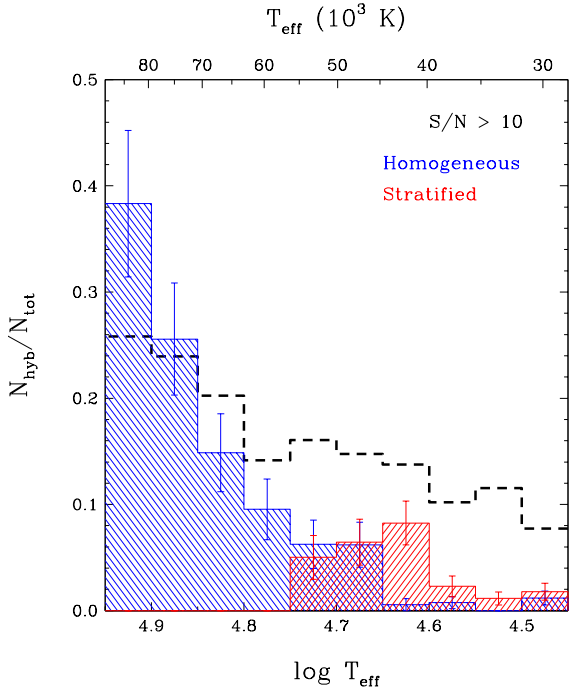}
\caption{Fraction of hybrid white dwarfs (corrected for the selection effects discussed in the text) as a function of effective temperature for our subsample of 1467 objects with S/N $\ge$ 10. The blue and red histograms correspond to stars with chemically homogeneous and stratified atmospheres, respectively. The error bars represent the Poisson statistics of each bin. For comparison purposes, the histogram of Figure \ref{fig:f_Teff_DO} is reproduced here as the black dashed line.}
\vspace{2mm}
\label{fig:f_Teff_hyb}
\end{figure}

Finally, we turn our attention to the hybrid white dwarfs. Such objects represent direct manifestations of the atmospheric metamorphoses occurring along the cooling sequence and are therefore of central interest for the theory of the spectral evolution. Since chemically homogeneous and stratified atmospheres arise from different physical mechanisms, we consider these two groups of stars separately. We show in Figure \ref{fig:f_Teff_hyb} the fractions of homogeneous and stratified white dwarfs in our sample as a function of decreasing effective temperature. The statistics are corrected for the selection effects described above, under the assumption (motivated by our findings) of thick and thin hydrogen layers for the homogeneous and stratified objects, respectively. Note that Figures \ref{fig:f_Teff_DO} and \ref{fig:f_Teff_hyb} use different horizontal and vertical scales; the histogram from the former is reproduced in the latter for ease of comparison.

The homogeneous white dwarfs, the bulk of which are DAO stars characterized by hydrogen-dominated atmospheres containing uniform traces of helium, exhibit a clear trend of declining incidence with decreasing effective temperature. They constitute more than 1/3 of the white dwarf population at $\Teff > 80,000$ K (although the shortcomings associated with this temperature range should be kept in mind), whereas they exist in very small numbers at $\Teff < 60,000$ K. The abrupt, continuous drop between the two regimes indicates that the physical process responsible for maintaining helium homogeneously in the outer hydrogen-rich layers rapidly becomes ineffective with cooling. The transport mechanism that best fits this description is a weak metal-driven stellar wind, as also alleged in previous studies of hot DAO white dwarfs \citep{bergeron1994,napiwotzki1999,gianninas2010}. Furthermore, other properties of these objects are consistent with the wind hypothesis: they typically have low masses and high metal contents (resulting in severe forms of the Balmer-line problem), two factors that stimulate mass loss and thereby helium contamination \citep{puls2000}. The idea of a steadily fading wind also implies a correlation between the helium abundance and the luminosity, as observed by \citet{napiwotzki1999} and predicted by \citet{unglaub2000}; we did not find such a correlation in our analysis, possibly because our ability to detect weak He {\sc ii} lines varies wildly from star to star, given the large range of S/N values among the SDSS spectra.

Anyhow, we reaffirm that residual stellar winds drive the spectral evolution of the hottest white dwarfs. At the beginning of the cooling sequence, trace species, that is, helium in hydrogen-rich objects as well as heavy elements in both hydrogen-rich and helium-rich objects, are supported uniformly in the outer envelope by this weak mass loss. Consequently, several members of the DA population (preferentially those with low masses and/or high primordial metal content) actually appear as homogeneous DAO stars. Since the strength of metal-driven winds depends mostly on the luminosity \citep{puls2000}, the cooling causes the wind to cease and thus the contaminants to settle: ultimately, homogeneous DAO stars become genuine DA stars. (Note that according to this interpretation, hybrid objects with homogeneous atmospheres should possess standard thick hydrogen layers, since they are the precursors of normal DA white dwarfs.) The calculations of \citet{unglaub2000} predict that this transition should happen at $\Teff \sim 85,000$ K for normal-mass objects. The fact that we observe a few homogeneous DAO stars down to lower effective temperatures implies that a small number of white dwarfs have stronger winds than expected, perhaps because they hold larger amounts of primordial heavy elements. Besides, we note that the occurrence of mass loss in extremely hot white dwarfs is in agreement with our previous claim that the sharp drop in the fraction of helium-dominated atmospheres at $\Teff \sim 90,000$ K seen in Figure \ref{fig:f_Teff_DO} is not due to some sort of spectral evolution. Indeed, gravitational settling does not operate in presence of a wind, hence it is highly implausible that a large number of objects experience a sudden DO-to-DA transformation through the float-up of residual hydrogen at such a high temperature. For this reason, we believe that our estimate of $\sim$24\% for the fraction of helium-rich stars at $\Teff \sim 75,000$ K is likely representative of the hotter white dwarf population as well.

The stratified white dwarfs are of a markedly different nature than their homogeneous counterparts. Figure \ref{fig:f_Teff_hyb} shows that they are much rarer, which explains why only one such object (PG 1305$-$017) was known prior to the SDSS \citep{bergeron1994,manseau2016}. They also tend to be cooler: they are all found in the effective temperature range where the fraction of helium-rich white dwarfs slowly decreases. Most importantly, the very fact that their atmospheres are chemically layered underlines that gravitational settling plays a dominant role here. Put together, these properties strongly suggest that stratified white dwarfs are transitional objects currently undergoing the DO-to-DA conversion, as advocated by \citet{manseau2016} and references therein. They are DO stars caught in the process of turning into DA stars through the upward diffusion of residual hydrogen, at that short-lived moment when the newly-formed superficial hydrogen layer is so thin that the underlying helium layer is still visible. With time, as more hydrogen initially diluted in the envelope will reach the surface, they will evolve into DA white dwarfs with thin hydrogen layers. 

The existence of objects with stratified atmospheres constitutes another clear proof that spectral evolution takes place among hot white dwarfs. Furthermore, these snapshots of the float-up process are found over a broad range of effective temperatures, which confirms that the temperature of the DO-to-DA transformation can differ significantly from star to star. On this topic, we note that all of our stratified white dwarfs have $\Teff < 55,000$ K, while we should also have discovered some with $75,000 \ {\rm K} > \Teff > 55,000$ K, given that we observe a decline in the fraction of helium-rich objects in this range as well. We believe that this non-detection is due to observational challenges rather than to a true absence. In fact, the spectroscopic distinction between homogeneous and stratified atmospheres becomes very subtle at high effective temperatures (see Figure 13 of \citealt{manseau2016}). Therefore, it is possible that some hot hybrid stars classified as homogeneous in this work will turn out to be stratified when observed at higher S/N. Furthermore, detecting a small amount of hydrogen in a hot helium-rich atmosphere is a difficult task even at high S/N, given that each H {\sc i} line is blended with a He {\sc ii} line (see Section \ref{sec:theory}). Yet another conceivable explanation would be that the DO-to-DA conversion occurs faster at higher temperatures, thus reducing our chances of catching it in the act.

\section{Summary and Conclusion} \label{sec:conclu}

We performed a comprehensive model-atmosphere analysis of 1806 hot ($\Teff \ge 30,000$ K) white dwarfs, including 1638 DA, 95 DO, and 73 DB stars, observed spectroscopically by the SDSS. We first presented new non-LTE model atmospheres and synthetic spectra computed with the codes TLUSTY and SYNSPEC, in which we implemented for the first time the detailed He {\sc i} line profiles of \citet{beauchamp1997}. We also introduced our next generation of evolutionary cooling sequences, which now extend to much higher effective temperatures and make use of more recent conductive opacities compared to our previous calculations. Our model atmospheres allowed us to analyze the SDSS optical spectra of all the objects in our sample and thus to homogeneously determine their atmospheric parameters, which were then converted into stellar parameters through our cooling tracks. 

At low effective temperature, most hydrogen-rich and helium-rich white dwarfs form a tight sequence around $M \sim 0.55$ \msun\ in the $\logg - \Teff$ diagram, with the exception of a few low-mass DA stars likely produced through common-envelope evolution. At high effective temperature, we found that the accuracy of the spectroscopic mass scale is compromised by significant systematic errors. On one hand, DA stars with $\Teff \ \gta 60,000$ K have lower-than-average masses, an effect that is possibly due to an improper calibration of the SDSS spectroscopic data as well as to the undetected presence of atmospheric metals in some objects. On the other hand, DO stars with $\Teff \ \gta 50,000$ K suffer from an opposite and even more severe problem, their masses being much too high, which we tentatively attributed to issues in the current modeling of He {\sc ii} line broadening.

Furthermore, we identified 127 white dwarfs exhibiting hybrid spectra, which we analyzed with two types of model atmospheres, assuming respectively homogeneous and stratified distributions of hydrogen and helium. We uncovered signs of atmospheric chemical stratification in 31 objects, thereby doubling the number of such white dwarfs known. We provided strong support to the idea that homogeneous and stratified objects constitute two fundamentally distinct groups, the former typically being considerably hotter and slightly less massive than the latter.

On the basis of our results, we were able to establish an exhaustive picture of the spectral evolution of hot white dwarfs, placing improved quantitative constraints on the theory of the spectral evolution. In particular, we are now in position to address the many outstanding questions raised in the Introduction:

1. What fraction of all white dwarfs are born with a helium atmosphere? At $\Teff \sim 75,000$ K, $\sim$24\% of white dwarfs are characterized by a helium-dominated surface composition. At the very highest effective temperatures ($\Teff > 90,000$ K), this proportion is markedly larger, but we believe this to be an artifact of the different cooling rates of extremely hot hydrogen-rich and helium-rich objects. This interpretation is consistent with the occurrence, at the very beginning of the cooling sequence, of weak stellar winds preventing any major changes in atmospheric composition. Our statistics at $\Teff \sim 75,000$ K are not affected by the above-mentioned evolutionary effect and are therefore more representative of the actual population of hot white dwarfs. In essence, our findings indicate that approximately one in four white dwarfs are born hydrogen-deficient.

2. Among those, how many eventually develop a hydrogen-rich atmosphere, and how many retain a helium-rich atmosphere throughout their life? The fraction of helium-dominated white dwarfs gradually decreases from $\sim$24\% at $\Teff \sim 75,000$ K to $\sim$8\% at $\Teff \sim 30,000$ K. From this result, it can be deduced that $\sim$2/3 of the DO stars turn into DA stars through the float-up of residual hydrogen before they cool down to $\Teff \sim 30,000$ K, and that the transformation takes place at different effective temperatures for different objects, depending on their total hydrogen content. As for the remaining $\sim$1/3, they most likely do not contain enough residual hydrogen for the DO-to-DA transition to happen and thus preserve a helium-dominated surface composition throughout their whole life. These two spectral evolution channels account for $\sim$16\% and $\sim$8\% of the total white dwarf population, respectively.

3. How does the number of hybrid white dwarfs, both with homogeneous and stratified atmospheres, vary with effective temperature? On one hand, homogeneous objects exist abundantly at $\Teff > 80,000$ K, where they represent more than 1/3 of our white dwarf sample. This proportion drops sharply with decreasing effective temperature, so that they become rather uncommon among cooler stars. On the other hand, stratified objects, which are intrinsically rarer than their homogeneous counterparts, concentrate mostly in the range $55,000 \ {\rm K} > \Teff > 40,000$ K, where they constitute $\sim$5$-$10\% of the white dwarf population, but a few of them are found down to $\Teff \sim 30,000$ K as well.

4. What does this imply about the role of stellar winds and the helium-to-hydrogen transition? In the case of the very hot DAO stars with homogeneous atmospheres, the trend uncovered in our study confirms that residual stellar winds govern the spectral appearance of the hottest white dwarfs. In particular, a weak mass loss initially causes small amounts of helium and heavier elements to be maintained uniformly in the external layers of hydrogen-rich objects, perhaps more efficiently in those having lower masses and/or higher metal contents. As cooling proceeds and winds accordingly fade, the trace species sink, thereby turning DAO stars into typical DA stars. For their part, the cooler white dwarfs showing stratified atmospheres are transitional objects in the process of undergoing the DO-to-DA conversion through the float-up of residual hydrogen. Their existence over a broad range of effective temperatures corroborates our assertion that the timing of this transformation varies substantially from one star to another.

5. What is the total hydrogen content of these various groups of white dwarfs, and how will it impact their future spectral evolution? The total masses of hydrogen characterizing the white dwarf population and giving rise to the observed spectral evolution cannot be precisely determined from our analysis alone. We may, however, speculate on the future spectral evolution specific to each channel identified here by connecting our findings to those obtained at lower effective temperatures ($\Teff < 30,000$ K). Approximately three in four white dwarfs are born as DA stars (or DAO stars if their wind is strong enough) and most likely possess canonical thick hydrogen layers ($q_{\rm H} \sim 10^{-4}$). Since no physical mechanism can alter such massive hydrogen layers along the cooling sequence, these objects will retain a hydrogen-dominated surface composition throughout their entire life \citep{tremblay2008,rolland2018,cunningham2020}. The remaining one in four white dwarfs are born as DO stars with much smaller amounts of hydrogen, which probably span many order of magnitudes. Most of them have enough hydrogen to transform into DA stars as a consequence of gravitational settling, but this state is only temporary. If the superficial hydrogen layer is sufficiently thin ($q_{\rm H} \ \lta 10^{-14}$), the convective dilution process will make the outer envelope helium-rich again before the star reaches $\Teff \sim 15,000$ K, and the spectral type will accordingly change to DB or DBA, and later on to DC \citep{macdonald1991,bergeron2011,koester2015,rolland2018,rolland2020,genest-beaulieu2019b}. Otherwise ($q_{\rm H} \ \gta 10^{-14}$), the white dwarf will still develop a helium-dominated atmosphere at a lower effective temperature via the convective mixing process, hence directly turning into a DC star or a helium-rich DA star \citep{macdonald1991,tremblay2008,chen2011,rolland2018,blouin2019,cunningham2020}. Finally, it is believed that those DO white dwarfs that are almost completely devoid of hydrogen will preserve essentially pure-helium atmospheres down to $\Teff \sim 10,000$ K, at which point they will evolve into DQ stars through the convective dredge-up of primordial carbon \citep{pelletier1986,dufour2005,koester2006,koester2019,koester2020,coutu2019}.

It is not an exaggeration to say that the spectral evolution of white dwarfs is now very well characterized from an empirical perspective, largely thanks to the advent of the SDSS and more recently of the {\it Gaia} mission. For this reason, we feel that the time is ripe for advances on the theoretical front, more specifically in the modeling of the dynamical phenomena invoked to explain the spectral evolution. This endeavor will require full-fledged evolutionary calculations, in which the appropriate chemical transport mechanisms (such as gravitational settling, convective mixing, and stellar winds, just to name a few) are self-consistently coupled to the cooling of a white dwarf. Only through such simulations will we gain the ability to answer the fundamental question at the center of spectral evolution theory: how much hydrogen is there in white dwarfs?

\acknowledgments 

This work was supported by the Natural Sciences and Engineering Research Council of Canada and the Fonds de Recherche du Qu\'ebec $-$ Nature et Technologie.

Funding for the Sloan Digital Sky Survey (\url{https://www.sdss.org}) has been provided by the Alfred P. Sloan Foundation, the U.S. Department of Energy Office of Science, and the Participating Institutions. SDSS-IV acknowledges support and resources from the Center for High-Performance Computing at the University of Utah, and is managed by the Astrophysical Research Consortium for the Participating Institutions of the SDSS Collaboration \citep{blanton2017}. 

This work has made use of data from the European Space Agency mission {\it Gaia} (\url{https://www.cosmos.esa.int/gaia}), processed by the {\it Gaia} Data Processing and Analysis Consortium (DPAC). Funding for the DPAC has been provided by national institutions, in particular the institutions participating in the {\it Gaia} Multilateral Agreement \citep{gaia2018}. 

This research has also made use of the NASA Astrophysics Data System Bibliographic Services; the Montreal White Dwarf Database \citep{dufour2017}; the SIMBAD database, operated at the Centre de Donn\'ees astronomiques de Strasbourg \citep{wenger2000}; and the NASA/IPAC Infrared Science Archive, operated at the California Institute of Technology.

\bibliographystyle{aasjournal}
\bibliography{main}

\clearpage
\startlongtable
\begin{longrotatetable}
\begin{deluxetable*}{llccccccccc}
\tabletypesize{\scriptsize}
\tablecolumns{12}
\tablewidth{0pt}
\tablecaption{Atmospheric and Stellar Parameters of Hot White Dwarfs from SDSS DR12 \label{tab:results}}
\tablehead{
\colhead{SDSS Name} &
\colhead{Type} &
\colhead{S/N} &
\colhead{$\Teff$ (K)} &
\colhead{$\logg$} &
\colhead{Composition} &
\colhead{$M/M_\odot$} &
\colhead{$R/R_\odot$} &
\colhead{log $L/L_\odot$} &
\colhead{log $\tau_{\rm cool}$} &
\colhead{Notes}
}
\startdata
 J000007.16$-$094339.9          & DA                             &   6.4                          &  30,469 (866)                  & 7.64 (0.22)                    & Pure H                         & 0.48 (0.09)                    & 0.0174 (0.0029)                & $-$0.63 (0.15)                 & 6.85 (0.17)                    &                               \\
 J000310.36$+$071801.1          & DA                             &  33.8                          &  30,532 (191)                  & 7.87 (0.05)                    & Pure H                         & 0.58 (0.02)                    & 0.0146 (0.0005)                & $-$0.78 (0.03)                 & 6.92 (0.01)                    &                               \\
 J000419.93$+$235000.2          & DA                             &  14.9                          &  38,163 (1447)                 & 7.70 (0.19)                    & Pure H                         & 0.52 (0.08)                    & 0.0169 (0.0025)                & $-$0.26 (0.15)                 & 6.59 (0.12)                    &                               \\
 J000509.93$+$003809.6          & DO                             &  35.2                          &  51,049 (564)                  & 8.17 (0.11)                    & Pure He                        & 0.74 (0.06)                    & 0.0117 (0.0009)                & $-$0.07 (0.07)                 & 6.28 (0.04)                    &                               \\
 J000622.61$+$010958.8          & DA                             &  29.6                          &  38,060 (578)                  & 7.67 (0.08)                    & Pure H                         & 0.51 (0.03)                    & 0.0174 (0.0010)                & $-$0.24 (0.06)                 & 6.58 (0.07)                    &                               \\
\enddata
\tablecomments{
\\ (1) The contribution of the main-sequence companion to the spectrum was removed before performing the fit.
\\ (2) At least one line core is filled with emission and was excluded from the fit.
\\ (3) The given parameters were obtained from chemically stratified model atmospheres, which yield a better fit to the hybrid spectrum.
\\ (4) The given parameters were obtained from chemically homogeneous model atmospheres, which yield a better fit to the hybrid spectrum.
\\ (5) The spectrum exhibits the Balmer-line problem and was fitted with the CNO-rich model atmospheres of \citet{gianninas2010}.
\\ (6) The traces of carbon were neglected in the model-atmosphere calculations, and the weak C {\sc iv} lines were excluded from the fit.
\\ (7) At least one line was excluded from the fit due to the presence of a glitch or to an incomplete wavelength coverage.
\\ (8) The main He {\sc ii} lines are abnormally strong and were excluded from the fit.
\\ (9) The given parameters are meaningless since the hybrid spectrum is actually produced by a DA+DO/DB binary system.
}
\end{deluxetable*}
\end{longrotatetable}

\end{document}